\documentclass[11pt]{article}
\usepackage[utf8]{inputenc}
\usepackage{amssymb}
\usepackage{lineno}
\usepackage{amsmath}
\usepackage{graphicx}
\usepackage{dsfont}
\usepackage{bm}
\usepackage{algorithm}
\usepackage{algpseudocode}
\usepackage{float}
\usepackage{xcolor}
\usepackage{subcaption}

\usepackage{amsmath, amssymb, amsthm, mathrsfs}
\usepackage[margin=1in]{geometry}
\usepackage{hyperref}
\hypersetup{
    colorlinks=true,
    linkcolor=blue,
    citecolor=red,
    urlcolor=cyan
}

\theoremstyle{plain}

\theoremstyle{definition}

\theoremstyle{remark}

\usepackage[utf8]{inputenc}
\usepackage{amsmath,amssymb,graphicx,geometry,hyperref,lineno}
\hypersetup{
    colorlinks=true,
    linkcolor=blue,
    citecolor=red,
    urlcolor=cyan
}

\begin{document}

\title{A Self-Exciting  Model of Eddy Formation at Submesoscale}
\author{Mine \c{C}a\u{g}lar \ and \ Bar{\i}\c{s} Samed Yakar \\Department of Mathematics, Ko\c{c}
University, Istanbul, Turkey}

\date{}

\maketitle

\abstract{
Ocean  eddies are highly dynamic structures marked by frequent splitting and merging events. They exhibit complex spatio-temporal clustering that traditional Poisson models fail to capture. In this paper,  a novel spatio-temporal Hawkes process is introduced to model self-exciting eddy fields on the basis of high-frequency ocean flow data. We formulate a triggering kernel that couples the self-excitation intensity with spatial deformation caused by the strain rate magnitude. To characterize the asymptotic behavior of the eddy field, we derive a Volterra  integral equation that governs the expected eddy intensity over time and space. We develop an Expectation-Maximization (EM) algorithm that treats the unobserved parent-child relationships as latent branching structures for parameter estimation. We further extend this framework to accommodate a time-varying, non-homogeneous background intensity, modifying both the EM updates and the analytical Volterra solution accordingly. Finally, we propose a two-stage simulation framework utilizing a cluster representation algorithm.
 The simulated empirical paths of eddy formation are compared together with numerical solution of the Volterra equation for mean rate as validation of the Hawkes model. 
}

\vspace{1mm}


\section{Introduction}
  
 Submesoscale eddies   play a critical role in ocean dynamics, coastal monitoring, and subgrid parameterizations for Large Eddy Simulations (LES) of ocean flows \cite{Kara2018}. To represent these small-to-medium flow structures, Çinlar random velocity field model has been successfully calibrated using high-frequency  radar observations along the Florida coast \cite{Caglar2006}. However,  Çinlar model assumes that eddy formation follows a Poisson process—a strong assumption that enforces independence between the appearances of distinct eddies over space and time. Oceanic eddy formation is inherently a clustered, spatio-temporal phenomenon. Observational and numerical studies demonstrate that larger eddies routinely split into smaller vortices, spawn secondary eddies, or merge \cite{Cui2019,Tian2026, Won2019, Zhang2022}. Empirically, submesoscale eddies occur in localized spatial clusters and temporal bursts rather than at memoryless, Poisson-distributed intervals. Capturing these self-exciting dynamics requires a point-process framework that explicitly accounts for event-to-event excitation. To address the limitations of the Poisson assumption, this study introduces a generalized spatio-temporal Hawkes process model tailored for submesoscale eddy formation by extending its preliminary form presented in \cite{Caglar2023}. By substituting a marked spatio-temporal Hawkes process for the traditional Poisson arrival process, our velocity field model captures the true temporal clustering and spatial triggering of ocean eddies.
 
The empirical foundation for our work relies on high-resolution Very High Frequency (VHF) surface-current radar observations of the Florida Current in a coastal domain of approximately \(11.25\,\mathrm{km}\times 11.25\,\mathrm{km}\).  Shay et al. \cite{Shay2000,Shay2002} reported very-high-frequency radar observations in this region with a spatial spacing of \(125\,\mathrm{m}\) and a temporal sampling interval of 15 minutes for 28 days. These measurements revealed episodic passages of submesoscale vortices with diameters of only a few kilometers. Such eddies are too small to be resolved directly by typical ocean general circulation models, motivating stochastic parameterizations for unresolved eddy-rich flows.
Subsequent studies examined this dataset from physical \cite{Peters2002, Shay2003}, modeling \cite{Mariano2003}, and statistical \cite{Caglar2006} perspectives. 
In  \c{C}inlar velocity field considered in \cite{Caglar2006}, the temporal dynamics follow a Poisson process where the event counts in disjoint regions are independent with an exponential interarrival distribution. On the other hand, the VHF-radar data indicate that eddies tend to occur in bursts  in time and in localized spatial regions.

In this paper,  we consider a spatio-temporal Hawkes model representing event-to-event excitation, in order to statistically capture the empirical behavior of eddy formation. 
A Hawkes process is a counting process $N(t)$ for the  number of eddies over time $t\ge 0$ with a random intensity function  \cite{Laub2021}.  Earlier environmental applications of Hawkes processes are  in seismology as a model for the progression of earthquakes and their aftershocks, as given in epidemic type aftershock sequence (ETAS) model \cite{Ogata1998,Reinhart2018}. As a new application, we leverage them  to develop precise estimation methods for eddy formation and triggering of new eddies in the ocean. The magnitudes of earthquakes serve as the marks of the Hawkes process in seismology whereas the marks are parameterized with the center, amplitude, radius and the lifetime of eddies  in the present paper. Although this type of modeling has a long history in seismology, the mathematical analysis given in \cite{Bielecki2022} is more recent in addition to earlier work of \cite{Daley2003}. 

We introduce the submesoscale velocity field  as an aggregation of eddies that emerge according to a Hawkes process $N(t)$ with intensity function $\lambda(t,z)$   indicating  the rate of occurrence of a new eddy at a specific time $t$ and location $z$. The dynamics of self-excitation is captured by a triggering function $g(t,z)$ contributing to $\lambda$ in addition to Poisson background arrivals. 
For temporal stationarity, Çinlar model incorporates a swift exponential decay of eddy magnitude whereas ocean data reveals a linear average trend instead. While a simple modification can resolve this structural discrepancy, the model's underlying assumption of a pure Poisson arrival process remains inadequate for capturing the true arrival dynamics. A Hawkes arrival process offers a superior alternative, maintaining stationarity as long as the so-called  the branching ratio  
\[
n=\mathbb{E}  \int_0^ \infty \! \! \int_{\mathbb{R}^2} g(t,z)\, dt \, dz
\]
is less than 1. The parameter $n$ represents the productivity factor, or the expected number of offsprings directly triggered by an arbitrary parent eddy. Because a Hawkes process overlays a self-exciting mechanism onto a baseline Poisson background, it improves model precision especially if a non-homogeneous Poisson process is used for the background arrival process. This framework fits our specific dataset remarkably well, yielding a notably small value for $n$ that indicates a subtle but distinct clustering behavior.

The main contributions of this work are three-fold:
\begin{enumerate}  
\item[] {\em Theoretical Framework:} We define a novel velocity field model driven by a Hawkes point process. To align with fluid dynamics, the spatio-temporal triggering kernel is modulated by local flow strain-rate parameters, directly linking event excitation to physical mechanisms. We also derive a Volterra integral equation for the expected intensity to analyze asymptotic behavior of the eddy field.
\item[] {\em Statistical Parameter Estimation:} Second, we develop an Expectation-Maximization (EM) algorithm tailored for statistical estimation of the  model parameters using eddy data. Treating the parent-offspring triggering relationships as latent branching variables ensures numerical stability and convergence.
\item[] {\em Simulation and Validation:} We present a dedicated simulation algorithm for Hawkes-driven velocity fields and validate parameter estimates against empirical VHF radar data and analytical Volterra solutions.
\end{enumerate}
 Consequently, the proposed velocity field provides a more accurate subgrid model for LES. Furthermore, these results are useful in risk assessment for tracking pollutant and hazardous material dispersion in coastal zones. All the estimation, simulation, and visualization codes are  available at
\href{https://github.com/barisyakar/Eddy-Estimation/}{https:/github.com/barisyakar/Eddy-Estimation/}. 

The paper is organized as follows. In Section \ref{2}, high-frequency radar data of the velocity field is described and analyzed for eddy statistics. The Hawkes type velocity field is developed theoretically in Section \ref{3}, with respect to its construction with a spatio-temporal triggering kernel as in ETAS model, the likelihood function for its later use in estimation, and the Volterra equation for the expected temporal intensity. In Section \ref{4}, the EM algorithm is proposed and applied to eddy data obtained after processing the raw data of HF radar measurements.  In Section \ref{5}, the estimation results are validated with simulation and with reference to Volterra solution for the expected intensity of eddy arrivals. According to these findings, the model fit is improved in Section \ref{6} by  allowing a background eddy arrival rate and modifying  the EM algorithm, updating the Volterra solution and the simulation method accordingly. Finally, the conclusions and future research directions are given in  Section \ref{7}.

\section{High-Frequency Data for Eddy-Rich Flows}
\label{2}
In  \cite{Caglar2006}, the raw data of vector fields have been processed to obtain eddy statistics such as arrival time, radius, center and magnitude of rotation.   In this section, we  provide supplementary analysis to motivate the new velocity model of the present paper.

\subsection{Eddy Statistics}
 
The observational setting of \cite{Shay2000} was used in \cite{Caglar2006} to construct an objective eddy-detection procedure to estimate eddy parameters such as center location, radius, amplitude, and lifetime. The resulting  statistics helped parameterizing a birth--death process of eddies through a stochastic velocity-field model of \c{C}inlar flows, which was later used as  a subgrid model in \cite{Kara2018} for LES. 

In the present paper, we use the processed eddy-event catalog as estimated in  \cite{Caglar2006} from the raw velocity vector field data obtained by radar. The   eddy-detection procedure and estimation of eddy parameters do not rely on the dynamic model of \c{C}inlar flows but they are obtained by fitting  Gaussian stream functions to vector field data at each snapshot of 15 minutes over 28 days. After the data are processed in this way, the sequence of variables obtained are denoted as 
$
t,\ l,\ x,\ y,\ a,\ b , 
$
where \(t\) has the form day.hour.minute, \(l\) is the estimated lifetime in minutes, \((x,y)\) are first-occurrence grid coordinates, \(a\) is the signed eddy amplitude, and \(b\) is the eddy radius. The coordinates are recorded on the \(125\,\mathrm{m}\) radar grid, and physical spatial distances are therefore obtained by multiplying grid-coordinate differences by \(125\,\mathrm{m}\). In the marked point-process model below, the formation time and location define the event, while \((a,b,l)\) are treated as marks.

\begin{figure}[h!]
    \centering
    \begin{subfigure}{0.49\textwidth}
        \centering
        \includegraphics[width=\linewidth]{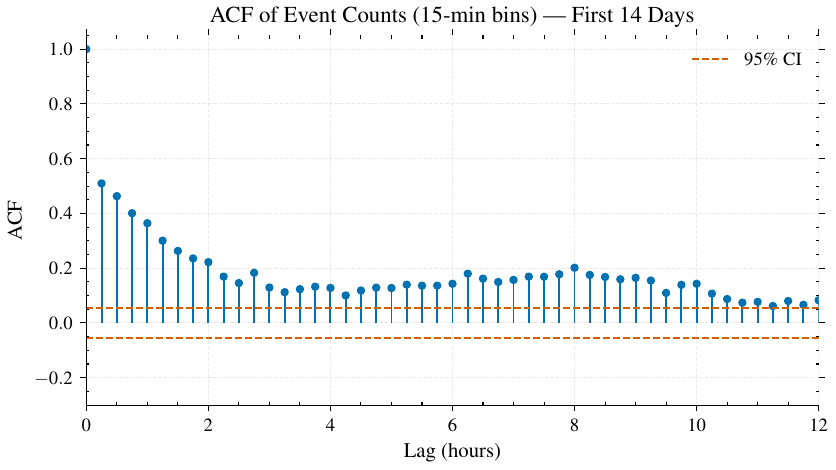}
        \caption{First 14 days.}
    \end{subfigure}
    \hfill
    \begin{subfigure}{0.49\textwidth}
        \centering
        \includegraphics[width=\linewidth]{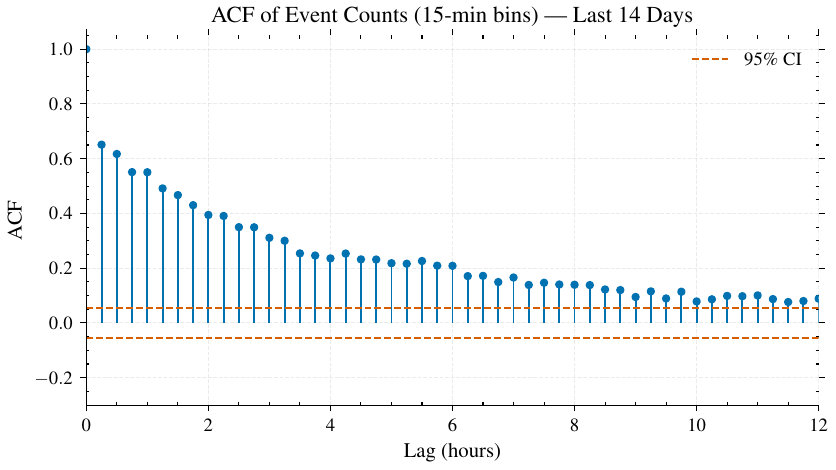}
        \caption{Last 14 days.}
    \end{subfigure}
    \caption{Autocorrelation function (ACF) of event counts at 15-minute resolution for the two observation windows. Positive short-lag autocorrelation indicates temporal clustering and departure from independent Poisson increments.}
    \label{fig:acf15}
\end{figure}

\begin{figure}[h!]
    \centering
    \begin{subfigure}{0.49\textwidth}
        \centering
        \includegraphics[width=\linewidth]{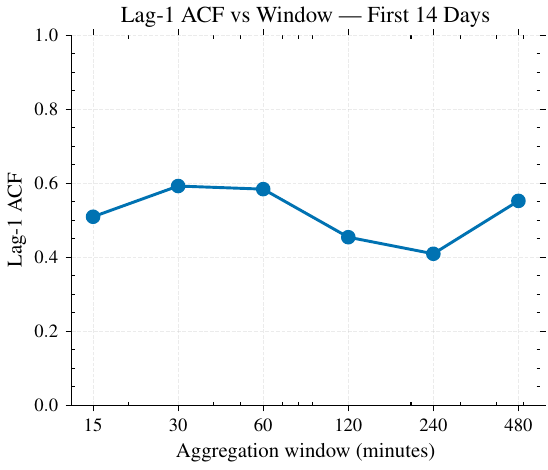}
        \caption{First 14 days.}
    \end{subfigure}
    \hfill
    \begin{subfigure}{0.49\textwidth}
        \centering
        \includegraphics[width=\linewidth]{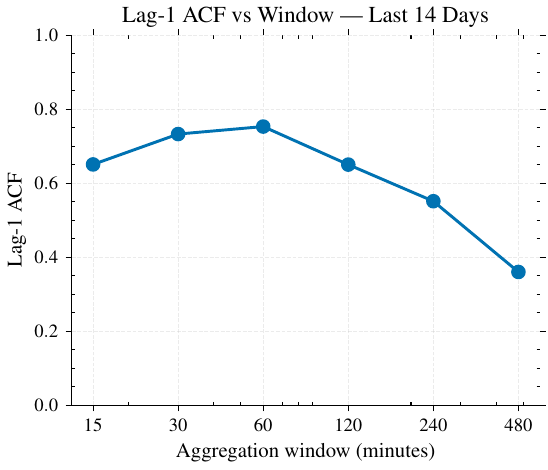}
        \caption{Last 14 days.}
    \end{subfigure}
    \caption{Lag-1 autocorrelation coefficient of event counts after aggregation over windows from 15 to 480 minutes. Persistence across aggregation scales provides evidence of temporal memory beyond the sampling interval.}
    \label{fig:acf_agg}
\end{figure}
 
\subsection{Temporal Dependence}

A defining characteristic of a homogeneous Poisson process is that its interarrival times are exponentially distributed, implying complete temporal independence. 
However, our data are sampled at discrete 15-minute intervals, causing the observed interarrival times to be integer multiples of 15 minutes. Since our data granularity is not high enough to test the goodness of fit of an exponential interarrival distribution, we compute the autocorrelation to show that the data reveal intrinsic temporal dependence that violates the Poisson assumption.

According to the  nonstationary behavior observed in  \cite{Caglar2006},   we divide the full observation period into two equal halves: the first 14 days and the last 14 days. To explore temporal structure, we compute the autocorrelation function (ACF) of event counts at 15-minute resolution and examine the evolution of lag-1 ACF across aggregated window sizes.
Let $C_k$ denote the number of eddy formations in the $k$-th 15-minute bin. We estimate temporal dependence using the sample autocorrelation function
\[
\hat r(h)=
\frac{\sum_{k=1}^{M-h}(C_k-\bar C)(C_{k+h}-\bar C)}
{\sum_{k=1}^{M}(C_k-\bar C)^2},
\]
where $M$ is the number of 15-minute bins in the corresponding 14-day window as plotted in Fig.~ \ref{fig:acf15}. The same binned count series is then aggregated over longer windows to examine whether lag-1 dependence persists beyond the original  sampling interval in Fig.~\ref{fig:acf_agg}.
The event counts exhibit substantial dependence, as evident from the positive autocorrelation values that persist even at large aggregation windows. This finding contradicts the independence assumption of Poisson processes. 
The persistence of positive autocorrelation across aggregation scales indicates that the temporal clustering is not only an artifact of the 15-minute sampling resolution. A Hawkes process, which is a self-exciting spatio-temporal model, is better suited to capture the observed dependence which leads to clustering dynamics.

\subsection{Spatial Distribution of Eddies}

\begin{figure}[h!]
    \centering
\begin{subfigure}{0.49\textwidth}
\includegraphics[width=\linewidth]{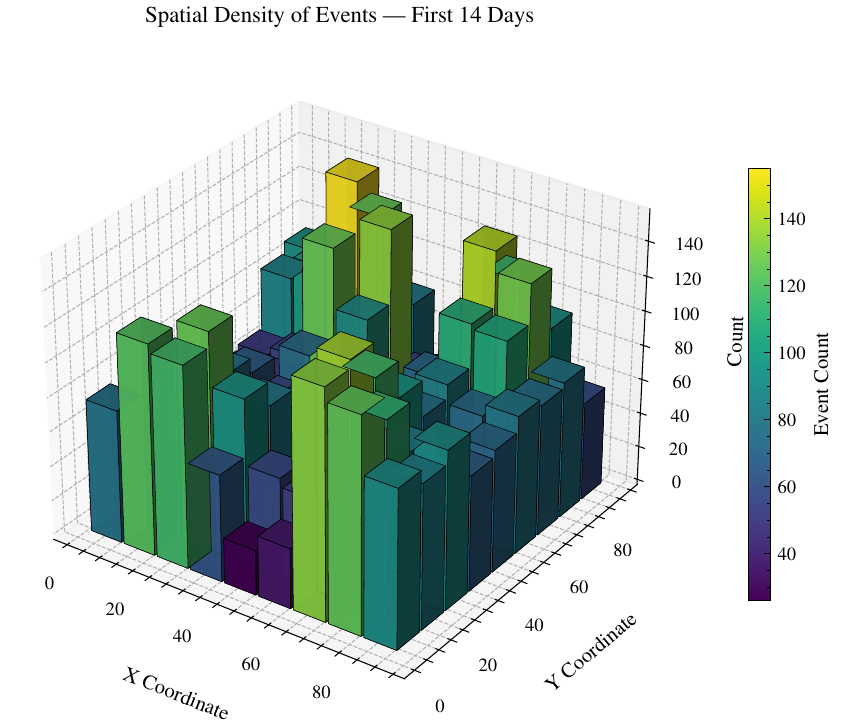}
       \caption{First 14 days}  
 \end{subfigure}
 \hfill
    \begin{subfigure}{0.49\textwidth}
    \includegraphics[width=\linewidth]{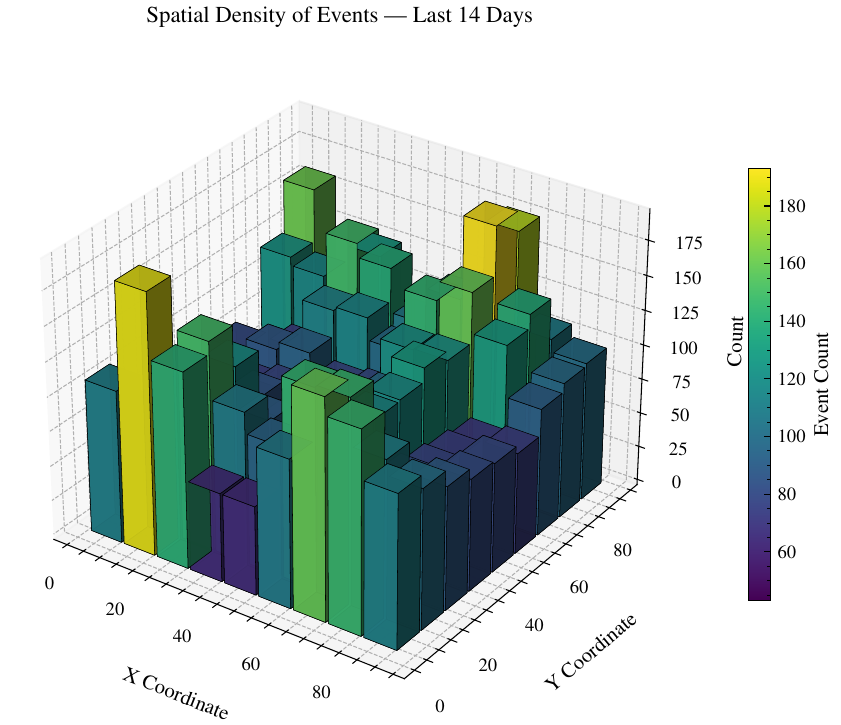}
  \caption{Last 14 days}
     \end{subfigure}
    \caption{Spatial density of eddies. The northern region ($Y > 70$) and southern region ($Y < 20$) show  pronounced hotspots, while central areas are relatively less active.}
    \label{fig:binned_spatial_map}
\end{figure}

\begin{figure}[h!]
    \centering
    \begin{subfigure}{0.49\textwidth}
        \centering
        \includegraphics[width=\linewidth]{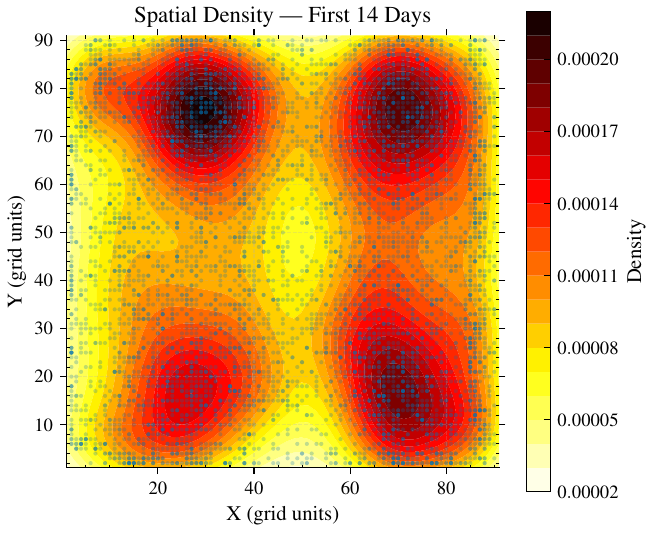}
        \caption{First 14 days}
    \end{subfigure}
    \hfill
    \begin{subfigure}{0.49\textwidth}
        \centering
        \includegraphics[width=\linewidth]{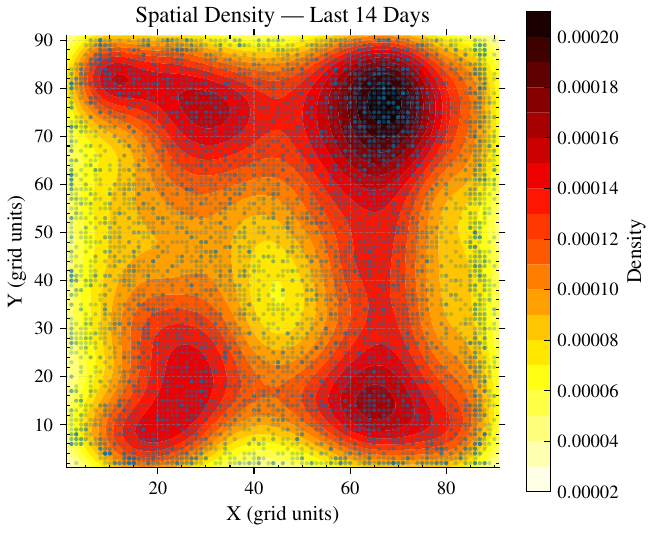}
        \caption{Last 14 days}
    \end{subfigure}
    \caption{Kernel-smoothed spatial density of eddy formation locations in the two observation windows. Both periods display strong spatial heterogeneity, with events concentrated in localized regions rather than uniformly distributed over the domain.}
    \label{fig:spatial_heatmap}
\end{figure}

The spatial density  given in Fig.~\ref{fig:binned_spatial_map}  shows higher number of eddies for $Y > 70$, and $Y<20$.  As kernel-smoothed version, 
Fig.~\ref{fig:spatial_heatmap}  shows a clear clustering pattern with a strong concentration of events in the northern region  whereas central and southern regions remain comparatively sparse. 
These patterns violate the spatial homogeneity assumption of \c{C}inlar model used in \cite{Caglar2006} and suggest the presence of local self-excitation mechanisms which we use in the Hawkes model. We take this empirical spatial distribution into account also in  simulations.

\begin{figure}[h!]
    \centering
    \includegraphics[width=0.72\textwidth]{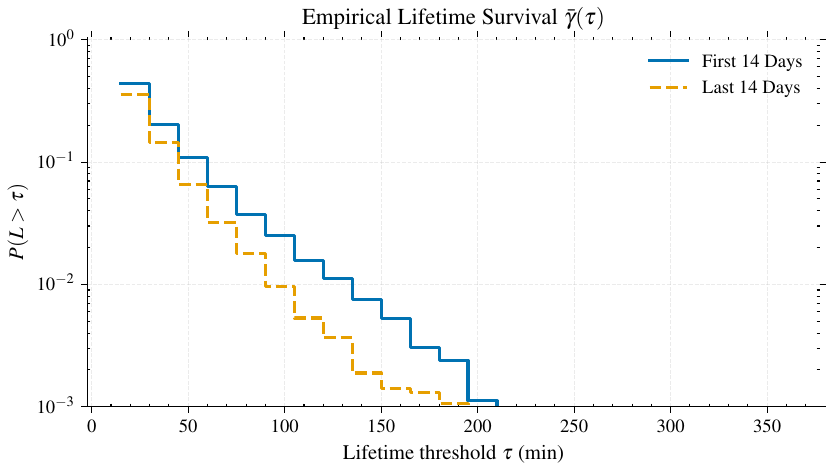}
    \caption{Empirical lifetime survival function $\bar{\gamma}(\tau)$ for the first and last 14-day windows.}
    \label{fig:lifetime_survival}
\end{figure}
In Fig.~\ref{fig:lifetime_survival}, we can conclude that  $\bar{\gamma}(\tau) \propto \exp{(-\alpha \tau)} $ for some $\alpha>0$ as the survival probability is in log-scale. 
Therefore, the lifetime of all eddies can be considered to follow an exponential distribution.   

\section{Model for Self-Exciting Eddies}
\label{3}
In this section, we construct a marked spatio-temporal Hawkes velocity model by developing the triggering kernel.

\subsection{Hawkes process and Hawkes driven velocity field} 
 
Let $N$ be a counting measure on the product space of Borel subsets of
$
\mathbb{R}_+ \times \mathbb{R}^2 \times \mathbb{R} \times  \mathbb{R}_+^2 
$
with coordinates $(t,z,a,b,l)$, where $t$ denotes time, $z=(x,y)\in \mathbb{R}^2$ represents spatial location and $(a,b,l)\in \mathbb{R}\times \mathbb{R}_+^2$ denotes a triplet of independent marks representing amplitude, radius, and lifetime of an eddy, respectively.
Let $(s_i,z_i,a_i,b_i,l_i)$, $i=1,2,\ldots$, denote the atoms of $N$. Then, the cumulative number of events up to time $t$ with centers in $A\subseteq \mathbb{R}^2$  and marks in $B\subseteq \mathbb{R}\times \mathbb{R}_+^2$  is
\[
N([0,t]\times A\times B)
= \sum_{i:s_i\le t} 1_A(z_i)1_B(a_i,b_i,l_i)
= \int_0^t \int_A \int_{B} N(ds,dz,da,db,dl).
\]
We take $N$ to be a Hawkes random measure with  compensator 
\begin{equation} \label{comp}
\Lambda(dt,dz,da,db,dl):=\lambda(t,z)\alpha(a)\beta(b){\gamma}(l)\,dt\,dz\, da\, db\, dl  
\end{equation}
in the sense that $N-\Lambda$ is a martingale \cite{Cinlar2011}, assuming that 
\begin{align}
    \lambda(t,z)
=&   \lambda_0 + \sum_{i:s_i<t} g(t,z| s_i,z_i,a_i,b_i,l_i) \nonumber  \\
=&   \lambda_0 + \int_0^{t^-} \int_{\mathbb{R}^2}\int_{\mathbb{R}_+^3}
 g(t,z|s,u,a,b, l)\, N(ds,du,da,db,dl)  
\label{eq:mark_intensity}
\end{align}
where $\lambda_0>0$ is the background rate per unit time-unit space, $\alpha,\beta,\gamma$ are the probability densities of the independent marks $a$,  $b$ and $l$, respectively, and $g$ is the triggering kernel. Instead of $g(t,z)$,   a more explicit notation $g(t,z|s,u,a,b, l)$  indicates the conditional intensity for the birth of a child eddy at time $t$ and location $z$ as trigerred by a parent eddy which has occurred at time $s$ with center $u$, amplitude $a$, radius $b$, and lifetime $l$.  

We construct the velocity field as an aggregation of eddies randomly arriving in time and space with stochastic parameters according to $N$. Let $N(t)$ denote the associated Hawkes process with the random measure $N$. That is, we use $N(t)$ instead of $N([0,t]\times\mathbb{R}^2 \times   \mathbb{R}\times \mathbb{R}_+^2) $, which counts the number of eddies that have arrived in $[0,t]$ over all space  with all possible marks. 
Let the velocity field  associated with a single eddy  of type $q := ({z}, a, b)$ be denoted by ${v}_q({x})$ as in \c{C}inlar velocity field. It is defined by scaling a deterministic ``basic eddy'' $ {v}$ as 
\begin{equation}
     {v}_q( {x}) = a  \, {v}\left( \frac{ {x} -  {z}}{b} \right)
    \label{eq:scaled_velocity}
\end{equation}
 at each $ {x}\in \mathbb{R}^2$. The basic eddy $ {v}( {\xi})$ is defined in normalized coordinates $ {\xi}: = \frac{( {x}- {z})}{b}$ as a rotation on the unit disk in $\mathbb{R}^2$ with  a shape function $m(\rho)$, where $\rho = | {\xi}|$ for $\rho\le 1$, as
  \begin{equation} \label{basic}
      {v}( {\xi}) = \frac{m(\rho)}{\rho} \begin{pmatrix} -\xi_2 \\ \xi_1 \end{pmatrix} 
 \end{equation}
 for $ {\xi}\in \mathbb{R}^2$.
In \cite{Caglar2000}, the compactly supported magnitude function 
\begin{equation} \label{magnitude}
m(\rho)=\begin{cases} \dfrac{1-\cos(2\pi\rho)}{2}, & 0\leq \rho\leq 1,\\[4pt] 0, & \rho>1 \end{cases} 
\end{equation}
is used for its smoothness in the derivation of analytical results and the value of $v$ at the origin is defined by continuity as $ {v}( {0})= {0}$. Hence, the basic eddy vanishes outside the unit disk, and the scaled eddy $ {v}_q$ is supported on $\| {x}- {z}\|\leq b$.

We define the submesoscale velocity field  by
\begin{equation}  \label{velocity}
 {u}( {x},t)=\sum_{i=1}^{N(t)} a_i (1-(t-t_i)/l_i)  {v}_{q_i}( {x}) \mathds{1}_{l_i > t - t_i}
\end{equation}
where $q_i = ( {z_i}, a_i, b_i)$ and the amplitude at time $t$ is taken to be a linearly decreasing  function from its initial
value $a_i$ until the eddy dies over the course of its specific lifetime $l_i$. 
For temporal stationarity, Çinlar model incorporates an exponential decay of eddy amplitude $a_i$ with $  \exp(-c(t-t_i))$, $c>0$,  at time $t$.   However, empirical ocean data reveals a linear average trend  as incorporated in the in the Hawkes velocity field \eqref{velocity} with the factor $(1-(t-t_i)/l_i) \mathds{1}_{l_i > t - t_i}$   \cite{Caglar2006}.

\subsection{Strain-modulated triggering kernel}

 Next, we define the triggering kernel $g$, tailored specifically for ocean eddy formation.
A central hypothesis of our model is that the generation of new turbulent structures is driven by the local deformation of the fluid. Previously, parameters such as the eddy arrival rate $\lambda$ have been modeled as proportional to the resolved strain rate since background strain rate shears, stretches, and splits eddies~\cite{Kara2018}. Consistent with the scaling arguments presented in \cite{Kara2018}, the instantaneous new eddy excitation of the process is taken to be proportional to the strain rate magnitude, that is, we assume 
\begin{equation}
    g \propto |S^{(q)}( {x})|.
\end{equation}
This formulation ensures that regions of high fluid deformation statistically exhibit a higher likelihood of triggering subsequent turbulent events and the triggering kernel $g$  reflects how the surrounding fluid flow actively deforms and destabilizes these structures.
To implement this approach, we require an explicit analytical expression for the magnitude of the strain rate tensor, $|{S}|$, induced by these individual eddies.

The strain rate tensor $S_{ij}$ is defined as the symmetric part of the velocity gradient tensor~\cite{Tennekes1972} by
\begin{equation*}
    S_{ij} = \frac{1}{2} \left( \frac{\partial u_i}{\partial x_j} + \frac{\partial u_j}{\partial x_i} \right).
\end{equation*}
Let us rewrite Eq.\eqref{basic} as
 \[
      {v}( {\xi}) = f(\rho) \begin{pmatrix} -\xi_2 \\ \xi_1 \end{pmatrix}
 \]
 where $ {\xi}=(\xi_1 , \xi_2 )$, and $f(\rho) = m(\rho)/\rho$, defined for brevity in subsequent derivations. To determine the contribution of a single eddy to the local strain, we apply the chain rule to Eq.~\eqref{eq:scaled_velocity}. The spatial derivatives in the physical domain scale inversely with the eddy radius $b$ as
\[
    \frac{\partial}{\partial x_j} = \frac{1}{b} \frac{\partial}{\partial \xi_j}.
\]
Consequently, the strain rate tensor for a generic eddy $q$ scales as
\[
    S_{ij}^{(q)}( {x}) = \frac{|a|}{b} S_{ij}^{\text{basic}}( {\xi}).
\]
We calculate the components of the basic strain rate tensor $S_{ij}^{\text{basic}}$ explicitly. For the shear component $S_{12}^{\text{basic}} = S_{21}^{\text{basic}}$, substituting the off-diagonal gradients yields:
\begin{equation*}
    S_{12}^{\text{basic}} = \frac{1}{2} \left[ \left( -f(\rho) - \frac{\xi_2^2}{\rho} f'(\rho) \right) + \left( f(\rho) + \frac{\xi_1^2}{\rho} f'(\rho) \right) \right]=\frac{1}{2} \frac{f'(\rho)}{\rho} \left( \xi_1^2 - \xi_2^2 \right).
\end{equation*}
The magnitude squared of the strain rate is given by $|S|^2 = 2 S_{ij} S_{ij}$. Substituting the derived components for normal and shear strains yields
\begin{align*}
    |S^{\text{basic}}|^2 &= 2 \left[ (S_{11})^2 + (S_{22})^2 + 2(S_{12})^2 \right] \\
     &= 2 \left[ \left(-\frac{\xi_1 \xi_2}{\rho} f'(\rho)\right)^2 + \left(\frac{\xi_1 \xi_2}{\rho} f'(\rho)\right)^2 + 2 \left( \frac{1}{2} \frac{f'(\rho)}{\rho} \left( \xi_1^2 - \xi_2^2 \right) \right)^2 \right] \\
     &= \rho^2 [f'(\rho)]^2.
\end{align*}
Transforming back to physical variables, where $\rho = r/b$ and $r = | {x} -  {z}|$, we obtain the closed-form expression for the strain rate magnitude squared induced by a single eddy as
\begin{equation}
    |S^{(q)}( {x})|^2 = \frac{a^2 r^2}{b^4} \left[ \frac{d}{d\rho}\left( \frac{m(\rho)}{\rho} \right) \bigg|_{\rho=r/b} \right]^2
    \label{eq:final_strain}
    = \frac{a^2}{b^2}  \rho^2 [f'(\rho)]^2= \frac{a^2}{b^2}|S^{\text{basic}}|^2
\end{equation}

\bigskip
Before integrating Eq.~\eqref{eq:final_strain} into the stochastic model, we verify its dimensional consistency and physical interpretation. This justification is crucial for validating the use of strain rate as a triggering covariate.

Let $L$ denote the dimension of length and $T$ the dimension of time. The parameters of the Çinlar eddy are defined as follows:
\begin{itemize}
    \item Amplitude $a$: Represents the characteristic velocity scale $[a] \sim L T^{-1}$.
    \item Radius $b$: Represents the spatial scale $[b] \sim L$.
    \item Distance $r$: Physical distance from the eddy center $[r] \sim L$.
\end{itemize}
The term inside the brackets of Eq.~\eqref{eq:final_strain} is the derivative with respect to a dimensionless variable $\rho = r/b$, rendering it dimensionless. Performing a dimensional analysis on the remaining coefficients yields:
\[
    [|S|^2] = \frac{[a]^2 [r]^2}{[b]^4} = \frac{(L^2 T^{-2}) (L^2)}{L^4} = T^{-2}.
\]
Thus, the derived magnitude $|S|$ correctly carries units of frequency ($T^{-1}$).

Physically, this result implies that the characteristic strain rate scales as $|S| \sim |a|/b$. This ratio represents the turnover frequency of the eddy. In the context of the turbulence energy cascade, this frequency governs the timescale of vortex stretching and breakup. Consequently, $|S|$ is the natural physical quantity to scale the rate of new eddy generation in our Hawkes model. 

We have the following model of $g$ given an eddy $i$ with arrival time $t_i$, center $z_i $, amplitude $a_i$, radius $b_i$, and lifetime $l_i$:
\begin{equation}
g(t,z \mid s_i,z_i,a_i,b_i,l_i)
= K_0 \frac{|a_i|}{b_i}\;\omega\, e^{-\omega(t - t_i)}\;\frac{\rho}{\pi}\, e^{-\rho \|z - z_i\|^2}\;\mathds{1}_{l_i > t - t_i},
\label{eq:trigger}
\end{equation}
where the parameter $K_0>0$ controls the excitation strength, $\omega > 0$ is the temporal decay rate, and $\rho > 0$ is the spatial decay rate.
Our choice of the triggering function \eqref{eq:trigger} is  spatio-temporal as in the ETAS triggering kernel, but with different decaying functions.  The temporal component is chosen to be an exponential function in relation to the empirical lifetime distribution  discussed in Fig.~\ref{fig:lifetime_survival}. The spatial decay function is taken as Gaussian, which is motivated by the idealized rotational form of the eddies. 
The contribution of the strain rate  magnitude is represented by the ratio $|a_i|/b_i$. The effect of eddy $i$ is valid only while it remains active as captured by the indicator term involving $l_i$.

\subsection{Likelihood function}
Let $\mathcal{D} = \{ (t_i, z_i, a_i, b_i, l_i) \}_{i=1}^{N(T)}$ denote the observed event catalog over time window $[0,T]$ . Because we do not observe the true branching structure , that is, whether an event is a background event or triggered by a specific prior event, we must formulate the likelihood based on the total conditional intensity of the process.
For a spatio-temporal Hawkes process, the total conditional intensity at an event $i$ is the sum of the background rate and the triggering contributions from all preceding events.
The data log-likelihood, $\ell(\Theta)$, for a point process observed over a spatial window $S\subset\mathbb{R}^2$ and time $[0,T]$ is given by
\begin{equation} \label{likelihood}
\ell(\Theta) = \sum_{i=1}^N \log\left(\lambda(t_i, z_i)\right) - \int_0^T \int_S \lambda(t, z) \,dz \,dt
\end{equation}
where we have the parameter vector $\Theta := (\lambda_0, K_0, \omega, \rho)$ from \eqref{eq:mark_intensity} and \eqref{eq:trigger}.
The general formulation of the log-likelihood for marked point processes is established by \cite{Daley2003}.

Substituting our specific intensity function into \eqref{likelihood}, the integral of the intensity separates into the background contribution and the expected number of offspring from all events. This yields the incomplete-data log-likelihood
\begin{equation}
\ell(\Theta) = \sum_{i=1}^N \log\left(\lambda_0 + \sum_{j: t_j < t_i} g(t_i, z_i \mid t_j, z_j, a_j, b_j, l_j)\right) - \left( \lambda_0 |S| T + \sum_{j=1}^N G_j  \right),
\label{eq:incomplete_likelihood}
\end{equation}
where $G_j$ is the expected number of offspring triggered by event $j$ over $S$, obtained by integrating the kernel \eqref{eq:trigger} over space and time as
\[
G_j=\int_0^T \int_S g(t,z \mid  t_j, z_j, a_j, b_j, l_j)
\]
due to the fact that the number of offspring triggered by event $j$ over $S$ is conditionally Poisson distributed with mean $G_j$ given event $j$.
Directly maximizing $\ell(\Theta)$ with respect to $\Theta$ is analytically intractable because the parameters appear inside the logarithm of a sum. This motivates the use of the EM algorithm, which iteratively maximizes a simpler likelihood function.

\subsection{Volterra  equation for mean intensity} \label{volterra_derivation}
 A Volterra integral equation for a function $u$ in its linear second-kind form, is written as $u(t)=f(t)+\int_0^t k(t-s)u(s)\,ds$ \cite{Gripenberg1990}. The function $k$ induces a causal structure, which  is natural for a Hawkes process because events formed before $t$ contribute to the current intensity through the triggering kernel. 

Taking expectation in \eqref{eq:mark_intensity} and using the compensation property of counting measures \cite[Prop.3.1]{Laub2021}, we obtain
\begin{eqnarray*}
\lambda^*(t,z)
&:= & \mathbb{E} [\lambda(t,z)] \\
&= &    \lambda_0
 + \mathbb{E}\left[\int_0^{t^-} \!\!\int_{\mathbb{R}^2}\int_{\mathbb{R}}\int_{\mathbb{R}_+^2}
 g(t,z \mid s,u,a,b, l)\, N(ds,du,da,db, dl)\right]    \\
 &= &    \lambda_0
 + \mathbb{E}\left[\int_0^t \int_{\mathbb{R}^2}\int_{\mathbb{R}} \int_{\mathbb{R}_+^2}
 g(t,z\mid s,u,a,b, l)\, \Lambda(ds,du,da,db, dl)\right]    \\
 &= &    \lambda_0
 + \mathbb{E}\left[\int_0^t \int_{\mathbb{R}^2}\int_{\mathbb{R}} \int_{\mathbb{R}_+^2}
 g(t,z\mid s,u,a,b, l)\,\alpha(a)\beta(b) {\gamma}(l)\lambda(s,u) ds\, du\, da\, db\,  dl\right]    \\
 &=&  \lambda_0
 +  \ \int_0^t \int_{\mathbb{R}^2}\int_{\mathbb{R}}\int_{\mathbb{R}_+^2}
 g(t,z \mid s,u,a,b, l)\,\alpha(a)\beta(b) {\gamma}(l)  \, \lambda^*(s,u) ds\, du\, da\, db\, dl  \\
  &=&  \lambda_0
 +  \ \int_0^t \int_{\mathbb{R}^2}\int_{\mathbb{R}} \int_{\mathbb{R}_+^2}
 K_0 \;\: \omega\, e^{-\omega(t - s)}\;\frac{\rho}{\pi}\, e^{-\rho \|z - u\|^2}\;\: \frac{|a|}{b}  \\
 & & \hspace*{6cm} \,\alpha(a)\beta(b) {\gamma}(l)\,\mathds{1}_{\{l>t-s\}} \lambda^*(s,u) ds\, du\, da\, db\,  dl.
\end{eqnarray*}
Integrating over the source marks $(a,b)$ yields the mean contribution of the strain rate through $|a|/b$, and  the integral over \(l\) with the indicator yields the survival function $\bar{\gamma}(\tau):=\int_{\tau}^{\infty}\gamma(l)\,dl=\mathbb{P}(L>\tau)$. Explicitly, we have
\begin{align*}
\int_{\mathbb{R}_+^3} \frac{|a|}{b}\,\mathds{1}_{\{l>t-s\}}\,\alpha(a)\beta(b)\gamma(l)\,da\,db\,dl
&= \left(\int_{\mathbb{R}_+^2}\frac{|a|}{b}\,\alpha(a)\beta(b)\,da\,db\right)
\left(\int_{t-s}^{\infty}\gamma(l)\,dl\right) \\
&= \mathbb{E}\!\left[\frac{|a|}{b}\right]\;\bar{\gamma}(t-s)\, .
\end{align*}
Hence, the equation simplifies to
\begin{equation}
\lambda^*(t,z)
= \lambda_0 + K_0\,\mathbb{E}\!\left[\frac{|a|}{b}\right]
\int_0^t \omega\, e^{-\omega(t-s)}\, \bar{\gamma}(t-s)\,
\left[\int_{\mathbb{R}^2} \frac{\rho}{\pi}\, e^{-\rho\|z-u\|^2}\,\lambda^*(s,u)\,du\right] ds.
\label{eq:expectedintensity}
\end{equation}

Let $S\subset\mathbb{R}^2$ be a finite spatial domain and define the spatially integrated expected intensity
\[
Q(t) = \int_S \lambda^*(t,z)\,dz.
\]
Integrating \eqref{eq:expectedintensity} over $S$ and exchanging the order of integration, we get
\begin{align*}
Q(t)
&= \lambda_0 |S|
 + K_0\,\mathbb{E}\!\left[\frac{|a|}{b}\right]
 \int_0^t \omega\, e^{-\omega(t-s)}\, \bar{\gamma}(t-s)\,
 \left[\int_{\mathbb{R}^2}\lambda^*(s,u)\, H_S(u)\,du\right] ds,
\end{align*}
where the spatial influence function is now:
\[
H_S(u) = \int_S \frac{\rho}{\pi}\, e^{-\rho\|z-u\|^2}\,dz.
\]
If $\lambda^*(s,u)$ is approximately spatially homogeneous over $S$, then $\lambda^*(s,u) \approx Q(s)/|S|$, yielding the Volterra integral equation of the second kind given by
\begin{equation}
Q(t)
= \lambda_0 |S| + K_0\,\mathbb{E}\!\left[\frac{|a|}{b}\right]\bar{H}_S
\int_0^t \omega\, e^{-\omega(t-s)}\, \bar{\gamma}(t-s)\,Q(s)\,ds,
\label{eq:volterra}
\end{equation}
where
\[
\bar{H}_S = \frac{1}{|S|}\int_{\mathbb{R}^2} H_S(u)\,du.
\]
The exponential density for the spatial distance implies that $\int_{\mathbb{R}^2}\frac{\rho}{\pi}e^{-\rho\|z-u\|^2}\,dz = 1$ for  $u\in \mathbb{R}$, so $\bar{H}_S \to 1$ when $S$ is sufficiently large.

The {\em branching ratio} associated with the expected intensity is
\[
n \;=\; K_0\,\mathbb{E}\!\left[\frac{|a|}{b}\right] \int_0^\infty \omega\, e^{-\omega \tau} \,\bar{\gamma}(\tau)\,d\tau.
\]
It follows that
\begin{equation}
\lim_{t\to\infty} Q(t) = \frac{\lambda_0 |S|}{1-n}, \qquad
\lim_{t\to\infty} \lambda^*(t,z) = \frac{\lambda_0}{1-n} \quad \forall\, z \in S,
\label{stationaryrate}
\end{equation}
when $n < 1$ \cite{Gripenberg1990}. The Hawkes process $N(t)$ is \emph{subcritical} if $n<1$, \emph{critical} if $n=1$, and \emph{supercritical} if $n>1$.
Although $n>1$ implies a  nonstationary  process which explodes as $t\to\infty$, Volterra equation \eqref{eq:volterra} is well-defined on any finite interval $[0,T]$, because the kernel
\[
K(\tau) = K_0\,\mathbb{E}\!\left[\frac{|a|}{b}\right]\bar{H}_S\,\omega\, e^{-\omega\tau}\,\bar{\gamma}(\tau)
\]
is continuous and integrable on $[0,T]$. Standard results on linear Volterra equations yield existence and uniqueness of a solution \cite{Gripenberg1990}.

\section{EM Algorithm for Parameter Estimation}
\label{4}
We employ an EM algorithm to estimate the parameter vector $\Theta = (\lambda_0, K_0, \omega, \rho)$. This approach relies on the branching structure representation of the Hawkes process, where the observed process is viewed as a superposition of a background Poisson process and clusters of triggered events.

\subsection{Branching Structure and Complete Data Likelihood}

The observed-data likelihood \eqref{eq:incomplete_likelihood} requires taking the logarithm of the total conditional intensity, which involves a sum over all past events. This log-of-sums is analytically intractable to maximize directly. To overcome this, we introduce latent variables to describe the unobserved branching structure and construct a complete-data likelihood function. Then, we use it to devise an EM algorithm as  in \cite{Schoenberg2008}. 

Following \cite{Schoenberg2008}, the complete-data log-likelihood $\ell_c(\Theta)$ requires three components. Let  $u_j$ be the branching structure defined as $u_j=0$ if $j$ is a background eddy and $u_j=i$ if it is triggered by eddy $i$, let $n$ be the number of background events, and let $m_i$ be the number of events triggered by $i$. If $u_i$ were observed, then we would write the {\it complete} likelihood by considering the number of background events, the number of triggered events, and the normalized space-time distribution of the offspring of each event regardless of being background or triggered by some other event. In that case, we could define $n = \sum_{i=1}^{N} \mathds{1}_{\{u_i=0\}}$ as the total number of background events and $m_i = \sum_{k=1}^{N} \mathds{1}_{\{u_k=i\}}$ as the total number of events triggered by eddy $i$, by writing $N$ for simplicity instead of $N(T)$.
In view of \eqref{eq:trigger}, the expectation $G_i$ appearing in \eqref{eq:incomplete_likelihood} is found explicitly as
\begin{align} 
G_i &= \int_{t_i}^{\infty} \int_{\mathbb{R}^2} g(t, z \mid t_i, z_i, a_i, b_i, l_i) \,dz \,dt \nonumber \\
&= K_0 \frac{|a_i|}{b_i} \left( \int_{t_i}^{t_i+l_i} \omega e^{-\omega(t - t_i)} \,dt \right) \left( \int_{\mathbb{R}^2} \frac{\rho}{\pi} e^{-\rho \|z - z_i\|^2} \,dz \right) \nonumber \\
&= K_0 \frac{|a_i|}{b_i} \left( 1 - e^{-\omega l_i} \right). \label{eq:expected_offspring}
\end{align}
Then, the normalized  density for the location of the triggered event in time and space at $(t_j,z_j)$ is the kernel \eqref{eq:trigger} with $i\equiv u_j$ divided by $G_{u_j}(\Theta)$ and its logarithm is found as
\begin{equation} \label{19}
\log \frac{g(t_j, z_j)}{G_{u_j}} = \log(\omega) - \omega(t_j - t_{u_j}) + \log(\rho) - \log(\pi) - \rho \|z_j - z_{u_j}\|^2 - \log(1 - e^{-\omega l_{u_j}}).
\end{equation}

We construct the complete likelihood by considering three factors:
\begin{enumerate}
    \item[] {\em Background Events:} The number of background events follows a Poisson distribution with mean $\lambda_0|S|T$. The log-likelihood of observing exactly $n$ background events is:
    \[ n \log(\lambda_0 |S| T) - \lambda_0 |S| T - \log(n!) \]
    \item[] {\em Offspring Counts:} The number of offspring triggered by eddy $i$ follows a conditionally Poisson distribution with expectation $G_i = K_0\frac{|a_i|}{b_i}(1-e^{-\omega l_i})$. The log-likelihood of observing $m_i$ offspring is:
    \[ m_i \log(G_i) - G_i - \log(m_i!) \]
    \item[] {\em Event Densities:} For the specific events that are triggered, we must account for the normalized space-time density of their locations relative to their parent event, as defined in \eqref{19}.
\end{enumerate}
Combining the components yields the complete-data log-likelihood
\begin{align}
\ell_c(\Theta) &= \left\{ -\log(n!) - \lambda_0 |S| T + n \log(\lambda_0 |S| T) \right\} \nonumber \\
&\quad + \sum_{i=1}^N \left\{ -\log(m_i!) - K_0 \frac{|a_i|}{b_i} (1 - e^{-\omega l_i}) + m_i \log\left(K_0 \frac{|a_i|}{b_i} (1 - e^{-\omega l_i})\right) \right\} \nonumber \\
&\quad + \sum_{i: u_i \neq 0} \left\{ \log(\omega) - \omega(t_i - t_{u_i}) + \log(\rho) - \log(\pi) - \rho \|z_i - z_{u_i}\|^2 - \log(1 - e^{-\omega l_{u_i}}) \right\}. \nonumber
\end{align}
Dropping constants with respect to $\Theta$ over the branching structure yields the simplified log-likelihood function is
\begin{align}
\mathcal{L}(\Theta) &= n \log(\lambda_0) - \lambda_0 |S| T \nonumber + \sum_{i=1}^N \left[ m_i \log\left(K_0 \frac{|a_i|}{b_i} \omega \frac{\rho}{\pi}\right) - K_0 \frac{|a_i|}{b_i} (1 - e^{-\omega l_i}) \right] \nonumber \\
&\quad \quad \quad \quad - \sum_{i: u_i \neq 0} \left[ \omega(t_i - t_{u_i}) + \rho \|z_i - z_{u_i}\|^2 \right]. \label{eq:lc_simp}
\end{align}

\subsection{Expectation Step (E-Step)}
Since the true branching structure $u_i$ is unobserved, we calculate the conditional expectation of the complete-data log-likelihood given the observed catalog $\mathcal{D}$ and the current parameter estimates $\Theta^{(k)}$. The expectations of the indicator functions correspond to the branching probabilities:
\begin{align*}
q_i^{(k)} &:= \mathbb{E}[\mathds{1}_{\{u_i=0\}} \mid \mathcal{D}, \Theta^{(k)}] = \mathbb{P}(u_i = 0 \mid \mathcal{D}, \Theta^{(k)}) = \frac{\lambda_0^{(k)}}{\lambda(t_i, z_i \mid \Theta^{(k)})}, \\
p_{ij}^{(k)} &:= \mathbb{E}[\mathds{1}_{\{u_i=j\}} \mid \mathcal{D}, \Theta^{(k)}] = \mathbb{P}(u_i = j \mid \mathcal{D}, \Theta^{(k)}) = \frac{g(t_i, z_i \mid t_j, z_j,  a_j, b_j, l_j,\Theta^{(k)})}{\lambda(t_i, z_i \mid \Theta^{(k)})},
\end{align*}
where the total conditional intensity at event $i$ is:
\[
\lambda(t_i, z_i \mid \Theta^{(k)}) = \lambda_0^{(k)} + \sum_{j:t_j<t_i} g(t_i, z_i \mid t_j, z_j,  a_j, b_j, l_j, \Theta^{(k)}).
\]
This formulation naturally ensures 
\begin{equation} \label{optimalprob}
q_i^{(k)} + \sum_{j:t_j<t_i} p_{ij}^{(k)} = 1\: .
\end{equation}

\subsection{Maximization Step (M-Step)}
At iteration $k$ of the EM algorithm, the E-step provides the expected branching probabilities $p_{ij}^{(k)}$, the expected number of background events $\hat{n}^{(k)}$, the expected number of triggered events $\hat{m}_i^{(k)}$ by  eddy $i$, and the total expected triggered events $\hat{L}^{(k)}=\sum_{i=1}^N \hat{m}_i^{(k) }$. We maximize the expected complete-data log-likelihood to update $\Theta$.
The expected complete-data log-likelihood is derived by replacing the unobserved latent variables—the background count $n$, the offspring counts $m_i$, and the branching assignments $u_i$—with their conditional expectations from the E-step. 

We evaluate the expectation of \eqref{eq:lc_simp} term by term. First, the true number of background events $n$ is defined as the sum of the indicators of being background for all events, that is, $\sum_{i=1}^N \mathds{1}_{\{u_i=0\}}$.
Taking expectation given the data and the current parameters yields the total expected background events as
\[ \mathbb{E} \,[\, \sum_{i=1}^N \mathds{1}_{\{u_i=0\}} \mid \mathcal{D}, \Theta^{(k)} ] = \sum_{i=1}^N q_i^{(k)},\] which simplifies the first term of the likelihood to $\log(\lambda_0)\sum_{i=1}^N q_i^{(k)} - \lambda_0 |S| T$. 
Next, the true number of offspring triggered by eddy $i$ is $m_i = \sum_{j:t_i<t_j} \mathds{1}_{\{u_j=i\}}$. Its expectation yields the mean number of offspring for eddy $i$, denoted as $\hat{m}_i^{(k)} = \sum_{j: t_i<t_j} p_{ji}^{(k)}$. 
Finally, the last term in \eqref{eq:lc_simp} evaluates the spatial and temporal decay densities for all triggered events. By rewriting the sum over triggered events as an explicit sum over all possible parent-child pairs $(j, i)$ with $t_j < t_i$, we apply the indicator function $\mathds{1}_{\{u_i=j\}}$. Taking the expectation replaces this indicator with the expected branching probability $p_{ij}^{(k)}$, yielding the final term $\sum_{i=1}^N \sum_{j < i} p_{ij}^{(k)} \left[ \omega(t_i - t_j) + \rho \|z_i - z_j\|^2 \right]$. Combining these expected components constructs the complete objective function to be maximized in the M-step.
By taking expectation of \eqref{eq:lc_simp}, our objective function becomes
\begin{align*} \mathbb{E}[\mathcal{L}(\Theta) \mid \mathcal{D}, \Theta^{(k)}] &= \log(\lambda_0)\sum_{i=1}^N q_i^{(k)} - \lambda_0 |S| T \nonumber \\
&\quad + \sum_{i=1}^N \left[ \Big(\sum_{j:t_i<t_j} p_{ji}^{(k)}\Big) \log\left(K_0 \frac{|a_i|}{b_i} \omega \frac{\rho}{\pi}\right) - K_0 \frac{|a_i|}{b_i} (1 - e^{-\omega l_i}) \right] \nonumber \\
&\quad - \sum_{i=1}^N \sum_{j:t_j < t_i} p_{ij}^{(k)} \left[ \omega(t_i - t_j) + \rho \|z_i - z_j\|^2 \right]
\end{align*}

The maximization of the objective function is performed as follows.\\ \\
{\em Update for $\lambda_0$:}
Taking the derivative with respect to $\lambda_0$ yields the standard estimator over the spatial window area $|S|$ and the time window $T$ given by
\begin{equation*}
\hat{\lambda}_0 = \frac{\sum_{i=1}^N q_i^{(k)}}{|S|\, T}.
\end{equation*}
The second derivative $-\sum q_i^{(k)} / \lambda_0^2$ is strictly negative, confirming a global maximum.\\ \\
{\em Shape Parameters ($\omega, \rho$):}
Following the partial information approach of \cite{Schoenberg2008}, we estimate the temporal and spatial decay parameters by isolating the terms related to the elapsed time and squared distance, ignoring the contribution of the expected number of offspring $G_j$. This approximation leverages the fact that the shape parameters are predominantly determined by the spatial and temporal spreads of locally clustered events, leading to stable, closed-form updates.
For the temporal decay $\omega$, isolating $\log(\omega)$ and $-\omega(t_i-t_j)$ terms and maximizing them with respect to $\omega$ gives
\begin{equation*} 
\hat{\omega}^{(k+1)} = \frac{\hat{L}^{(k)}}{\sum_{i=1}^N \sum_{j < i} p_{ij}^{(k)} (t_i - t_j)}
\end{equation*}
where recall that $\hat{L}^{(k)}=\sum_{i=1}^N \hat{m}_i^{(k) }$. For the spatial decay parameter $\rho$, isolating $\log(\rho)$ and $-\rho\|z_i - z_j\|^2$ terms yields the estimator
\begin{equation*}
\hat{\rho}^{(k+1)} = \frac{\hat{L}^{(k)}}{\sum_{i=1}^N \sum_{j < i} p_{ij}^{(k)} \|z_i - z_j\|^2} \:.
\end{equation*}
\\ 
{\em Productivity Parameter $K_0$:}
Conditioning on the updated temporal shape parameter $\hat{\omega}^{(k+1)}$, we take the derivative of the expected complete-data log-likelihood with respect to $K_0$ and set it to zero
\begin{equation*}
\frac{\partial}{\partial K_0} \mathbb{E}[\mathcal{L}(\Theta)] = \frac{\hat{L}^{(k)}}{K_0} - \sum_{i=1}^N \frac{|a_i|}{b_i} (1 - e^{-\hat{\omega}^{(k+1)} l_i}) = 0.
\end{equation*}
Solving for $K_0$ yields the closed-form update:
\begin{equation*}
\hat{K}_0 = \frac{\hat{L}^{(k)}}{\sum_{i=1}^N \frac{|a_i|}{b_i} \left(1 - e^{-\hat{\omega}^{(k+1)} l_i}\right)}.
\end{equation*}

\subsection{EM Estimation Results}

Since the event catalog is observed only at 15-minute resolution, the occurrence times are effectively interval-censored rather than continuously observed. For that reason, the estimation procedure is carried out on discrete time blocks of width $\Delta = 15$ minutes. In particular, the temporal lags entering the E-step and M-step are computed in block units, which is the natural scale supported by the data. This avoids assigning artificial precision to inter-event times below the sampling resolution.

The constant-background Hawkes model is fitted separately to the first and
last 14-day observation windows. To maintain physical interpretability and
numerical stability, the random initializations were constrained as
\[
\lambda_0 \in [0.01\kappa, \kappa], \quad
K_0 \in [0.01, 10^6], \quad
\omega, \rho \in [0.1, 2],
\]
where $\kappa$ denotes the empirical mean arrival rate per unit area and per
time step. The observed catalogs contain 6269 and 8480 eddy arrivals in the first
and last 14-day periods, respectively. Since each period contains
\(T=1344\) 15-minute observation intervals, the corresponding
empirical mean arrival rates are \(\kappa=5.63\times10^{-4}\) and \(\kappa=7.62\times10^{-4}\) eddies per
15-minute interval per unit area for the first and second 14-day windows, respectively. Thus, the mean arrival
rate increases by approximately \(35.27\%\) in the second period. The upper bound $\lambda_0=\kappa$ corresponds to the limiting case
in which all events are attributed to the background process, while smaller
values allow part of the observed catalog to be explained by self-excitation.
The productivity parameter $K_0$ is initialized over a wide logarithmic range,
and the lower bounds on $\omega$ and $\rho$ enforce integrability of the
exponential temporal and Gaussian spatial kernels.

The resulting constant-background estimates are given in Table \ref{tab:constant_background_em_results}. Running the EM algorithm yields stable interior solutions for both observation windows. All random initializations converge to  unique final estimates of the parameters for both periods. 
The last 14-day window has a larger estimated background rate $\lambda_0$ but a smaller
productivity parameter $K_0$ than the first 14-day window. 
Therefore, the higher arrival rate $\kappa$ 
during the last 14 days can be attributed to a stronger exogenous
background component $\lambda_0$ in spite of a decreased self-excitation.

\begin{table}[htbp]
\caption{ EM estimates for the constant-background rate}
\label{tab:constant_background_em_results}
\centering
\small
\begin{tabular}{lcccc}
\hline
Period & $\hat{\lambda}_0$ & $\hat{K}_0$ & $\hat{\omega}$ & $\hat{\rho}$ \\
\hline
First 14 days & $4.9833\times 10^{-4}$ & $2.1154$ & $0.77657$ & $0.020512$ \\
Last 14 days  & $7.1145\times 10^{-4}$ & $0.79015$ & $0.83823$ & $0.019279$ \\
\hline
\end{tabular}
\end{table}

\section{Simulation of Spatio-Temporal Hawkes Process and Validation}
\label{5}
To validate the Expectation-Maximization estimation procedure and study the generative dynamics of eddy formation, we will simulate synthetic event catalogs from a known parameter vector $\Theta = (\lambda_0, K_0, \omega, \rho)$. 
Simulating a Hawkes process chronologically via thinning algorithms requires evaluating the conditional intensity $\lambda(t, z)$ continuously and rejecting candidate events. Because our spatial and temporal kernels are strictly defined by exponential and Gaussian decay, we can instead employ the much more efficient, rejection-free branching process representation \cite{HawkesOakes1974}. 

\subsection{Background Rate \texorpdfstring{$\lambda_0$}{lambda0}}

In the EM estimation, the background rate $\lambda_0$ is modeled as a \emph{scalar constant} representing the mean background intensity per unit time per unit area  and  estimated by
\[
\hat{\lambda}_0 = \frac{\sum_{i=1}^N q_i^{(k)}}{|S| \cdot T}, 
\]
where $k$ is the iteration at which the algorithm converges. The product
\[
\mu := \lambda_0 \cdot |S|
\]
represents the expected number of exogenously driven background  eddy formations \emph{per unit time} over the entire spatial domain $S$. This quantity provides a transparent physical baseline whereas the magnitude of endogenous self-excitation is quantified through the branching ratio $n$ and  the stationary expected total rate will be $\mu / (1 - n)$  in view of \eqref{stationaryrate}. The estimate of $n$ is obtained by
\[
\hat{n} \;=\; \hat{K}_0\, \overline{\frac{|a|}{b}} \int_0^\infty \hat{\omega}\, e^{-\hat{\omega} \tau} \,\bar{\gamma}(\tau)\,d\tau, 
\]
where $\overline{{|a|}/{b}}$ denotes the sample mean of $a_i/b_i$, $i=1,\ldots,N(T)$. Assuming a spatially uniform $\lambda_0$ in estimation is motivated by three considerations:
\begin{itemize}
    \item[] {\em Mathematical tractability.} A constant $\lambda_0$ yields closed-form M-step updates for all parameters and preserves the scalar Volterra integral equation structure~\eqref{eq:volterra}. Introducing a spatially varying $\lambda_0(x,y)$ would require a spatially-resolved Volterra system, substantially complicating the analysis. 

    \item[] {\em Parameter identifiability.} A nonparametric spatially varying background creates an identifiability tension with the spatial triggering kernel $(\rho/\pi)\exp(-\rho\|z - z_i\|^2)$. The EM algorithm could attribute spatial clustering to either the background density or the triggering structure, making the decomposition of self-excitation versus background ambiguous.

    \item[] {\em Physical interpretability.} The scalar $\mu = \lambda_0 \cdot |S|$ is the total exogenous eddy formation rate   separated from the endogenous, in other words triggered contribution.
\end{itemize}

\subsection{Spatial Distribution of Eddies}

Although $\lambda_0$ is estimated as a scalar for the reasons above, the \emph{empirical spatial distribution} of eddy formations is markedly non-uniform (cf.\ Fig.~\ref{fig:spatial_heatmap}). Generating background events uniformly over $S$ would therefore produce synthetic catalogs whose spatial structure is unrealistic. To reconcile this without altering the estimation model, we separate the \emph{rate} and \emph{spatial distribution} of background events via a two-stage approach. Here, we assume that the background eddies follow the spatial distribution of all eddies in the data set. This is justified also by our modeling of the triggered events as appearing in the vicinity of the background events. We simulate the locations of the eddies in two stages as follows.

\begin{itemize}
    \item[] {\em Offline density estimation:}
We estimate the spatial density of eddy formations from the empirical catalog using a two-dimensional Gaussian kernel density estimator. Let $\{(x_i, y_i)\}_{i=1}^N$ be the observed locations. 
The density estimate is
\begin{equation}
\hat{f}(x, y) = \frac{1}{N} \sum_{i=1}^N \frac{1}{2\pi h^2} \exp\!\left( -\frac{(x - x_i)^2 + (y - y_i)^2}{2h^2} \right),
\label{eq:spatial_kde}
\end{equation}
where $h = N^{-1/6}\,\hat{\sigma}$ is inspired by Silverman's rule for bivariate data \cite{Silverman2018}, with $\hat{\sigma} = \sqrt{(\hat{\sigma}_x^2 + \hat{\sigma}_y^2)/2}$.

\item[] {\em Spatially informed simulation:}
The total background count is drawn as $N_{\mathrm{b}} \sim \mathrm{Poisson}(\lambda_0 \cdot |S| \cdot T)$ using the scalar $\lambda_0$ from the EM estimation. However, the spatial locations of these events are sampled from the learned density~$\hat{f}$ rather than uniformly:
\[
(x_k, y_k) \sim \hat{f}, \quad k = 1, \ldots, N_{\mathrm{b}}.
\]
 This acts as a posterior calibration for the spatial EM estimator $\lambda_0$. The EM estimator $\hat{\lambda}_0 = \sum q_i^{(k)} / (|S|\,T)$ integrates over the entire domain and is therefore invariant to the spatial distribution of background events within~$S$. On the other hand, the triggering parameters $(K_0, \omega, \rho)$ depend on pairwise temporal delays and spatial distances, not on the marginal spatial density.

\end{itemize}

\subsection{The Cluster Representation Algorithm}

A Hawkes process admits an immigration--birth, or Poisson-cluster,
representation in which background events act as immigrants and every event
independently generates a Poisson process of direct offspring
\cite{HawkesOakes1974}. Rather than stepping through time, we simulate the catalog generation by generation.

First, the background events, also called immigrants or generation-zero
events, are generated from a Poisson process over $S\times[0,T)$.
The process is homogeneous in time but spatially inhomogeneous, with
background intensity
\[
\mu(z)=\lambda_0 |S|\hat f(z), \qquad z\in S,
\]
where $\hat f$ is the spatial density estimated from the observed eddy
locations. Consequently, the total number of background events satisfies
\[
N_{\mathrm b}\sim
\operatorname{Poisson}\!\left(\lambda_0|S|T\right),
\]
their occurrence times are uniformly distributed over $[0,T)$, and their
locations are independently sampled from $\hat f$.
 
Next, for each event in a given generation, we determine the number of direct offspring it triggers. In our model, the triggering capacity of a parent event $i$ depends on its amplitude $a_i$, scale $b_i$, and active lifetime $l_i$. The expected number of offspring $m_i$ is found by integrating the parent's triggering kernel over its lifetime as in \eqref{eq:expected_offspring}, so $m_i = G_i(\Theta)$. The exact number of offspring $N_i$ triggered by event $i$ is drawn from a Poisson distribution with mean $m_i$. 

For each triggered offspring, its relative time delay is drawn from a truncated exponential distribution defined by $\omega$ (bounded by the parent's lifetime $l_i$). Its spatial location is sampled directly from the spatial kernel $\frac{\rho}{\pi} e^{-\rho \|z - z_i\|^2}$, which is equivalent to a bivariate normal distribution centered at the parent's location $z_i$ with covariance matrix $\Sigma = \frac{1}{2\rho}\mathbf{I}_2$.

This recursive branching continues until a generation produces zero offspring within the observation window $T$. The complete generative procedure is summarized in Algorithm~\ref{alg:simulation_hawkes_cluster} in Appendix. It is based on the immigration--birth representation of \cite{HawkesOakes1974}. It is adapted
here to incorporate a two-dimensional spatial triggering kernel,
mark-dependent offspring productivity, event-specific eddy lifetimes,
empirically sampled eddy characteristics, and a KDE-based spatial background
distribution.

\subsection{Comparison with Data}

We validate the constant-background estimates in two complementary ways. First, the expected spatially integrated intensity is derived from the fitted Hawkes model through a Volterra equation, which provides a deterministic benchmark for the mean arrival rate. Second, synthetic catalogs are generated from the same fitted parameters using the branching representation, and their empirical arrival rates are compared with both the Volterra solution and the observed data. 

\begin{figure}[H]
    \centering
    \begin{subfigure}{\textwidth}
        \centering
        \includegraphics[width=\textwidth]{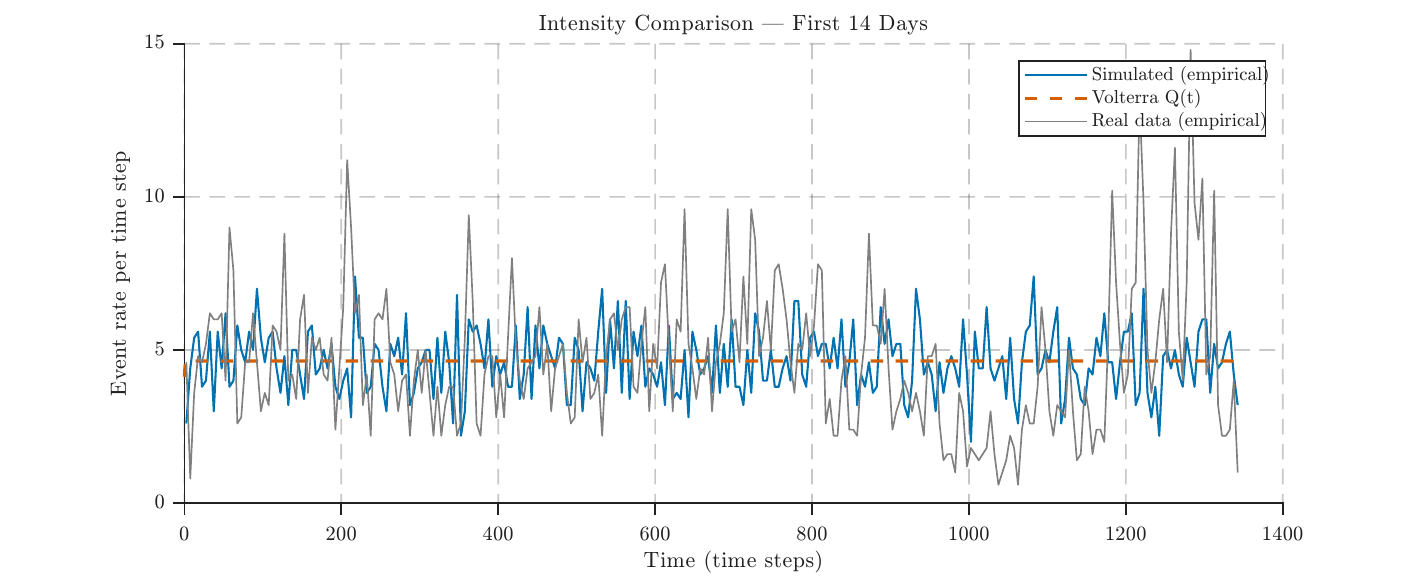}
        \caption{First 14 days.}
    \end{subfigure}

    \vspace{0.5em}

    \begin{subfigure}{\textwidth}
        \centering
        \includegraphics[width=\textwidth]{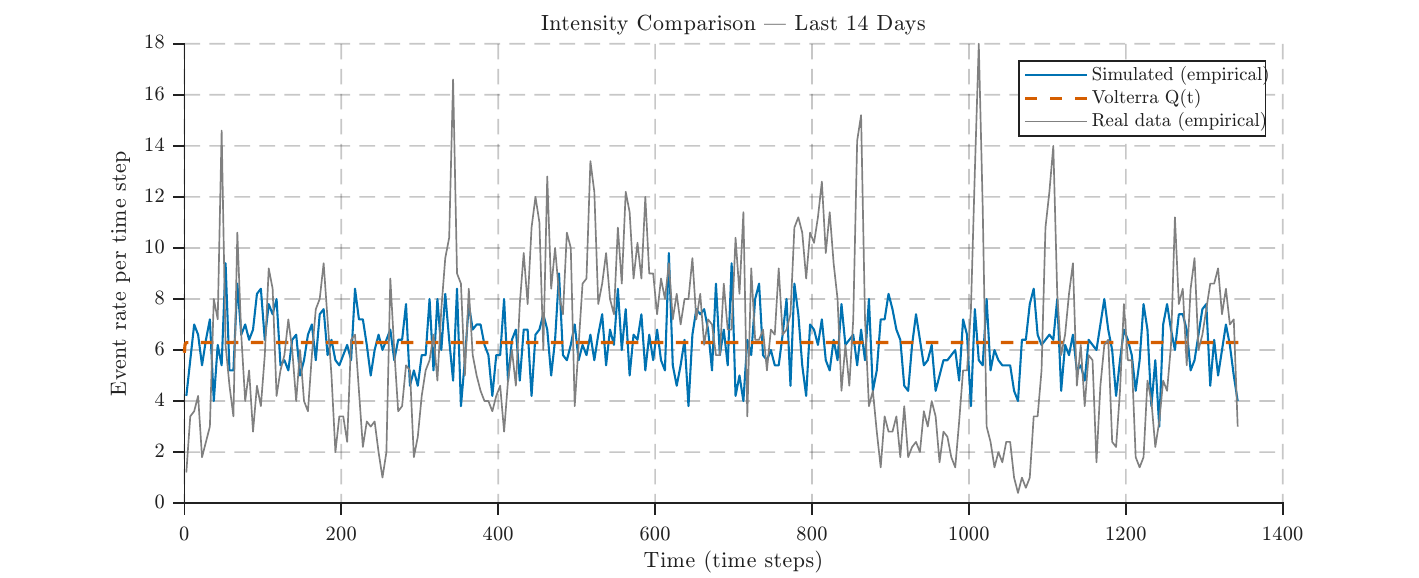}
        \caption{Last 14 days.}
    \end{subfigure}
    \caption{Constant-background simulation validation. The simulated empirical rate fluctuates around the expected rate found from Volterra equation, while the observed empirical rate exhibits higher variability in time. This discrepancy motivates the time-varying background model.}
    \label{fig:constant_background_intensity_comparison}
\end{figure}

Fig.~\ref{fig:constant_background_intensity_comparison} compares the empirical
arrival rate from one constant-background simulation with the Volterra expected
rate and the observed empirical arrival rate. In both 14-day windows, the
Volterra curve is flat throughout as  $\lambda_0$ is  constant. The simulated catalogs fluctuate around this constant
level, but they do not reproduce the broader low-frequency variation and burst
structure seen in the real eddy arrivals. This shows that the constant
background component does not contribute to the variance of the arrival-rate dynamics: it is
mathematically consistent with the fitted Hawkes model, but it is not flexible
enough to represent the changing exogenous conditions in the observed eddy
field. This validation separates correctness of the fitted Hawkes mechanism from its ability to reproduce the full nonstationary dynamics of the eddy catalog.
The dominant mismatch appears in the temporal
variability of the arrivals rather than in the local branching mechanism.

The simulated Hawkes catalog also provides the information needed to reconstruct
a stochastic velocity field at each observation time. For a fixed snapshot
\(t\), only eddies that are alive at that time are included. That is, eddy $i$ is considered only if $
t_i \leq t < t_i + l_i$.
Because the radar data are recorded at 15-minute resolution, we use the same
time step for the reconstruction. Thus, the first three days of a simulated
catalog correspond to \(3\times 24\times 4=288\) velocity-field snapshots.
On the spatial grid, the velocity vector $ {v}_q$ for an eddy are found using equations \eqref{eq:scaled_velocity},\eqref{basic}, and
  \eqref{magnitude}. Note that a Gaussian velocity profile could be used instead to obtain $ {v}_q$ by respecting the conversion of  parameters as in \cite[Sec.4]{Caglar2006}.
The full simulated velocity field is computed by superposition of all active
eddies as in Eq.~\eqref{velocity}.

\begin{figure}[htbp]
    \centering

    \begin{subfigure}[t]{0.485\textwidth}
        \centering
        \includegraphics[width=\linewidth]{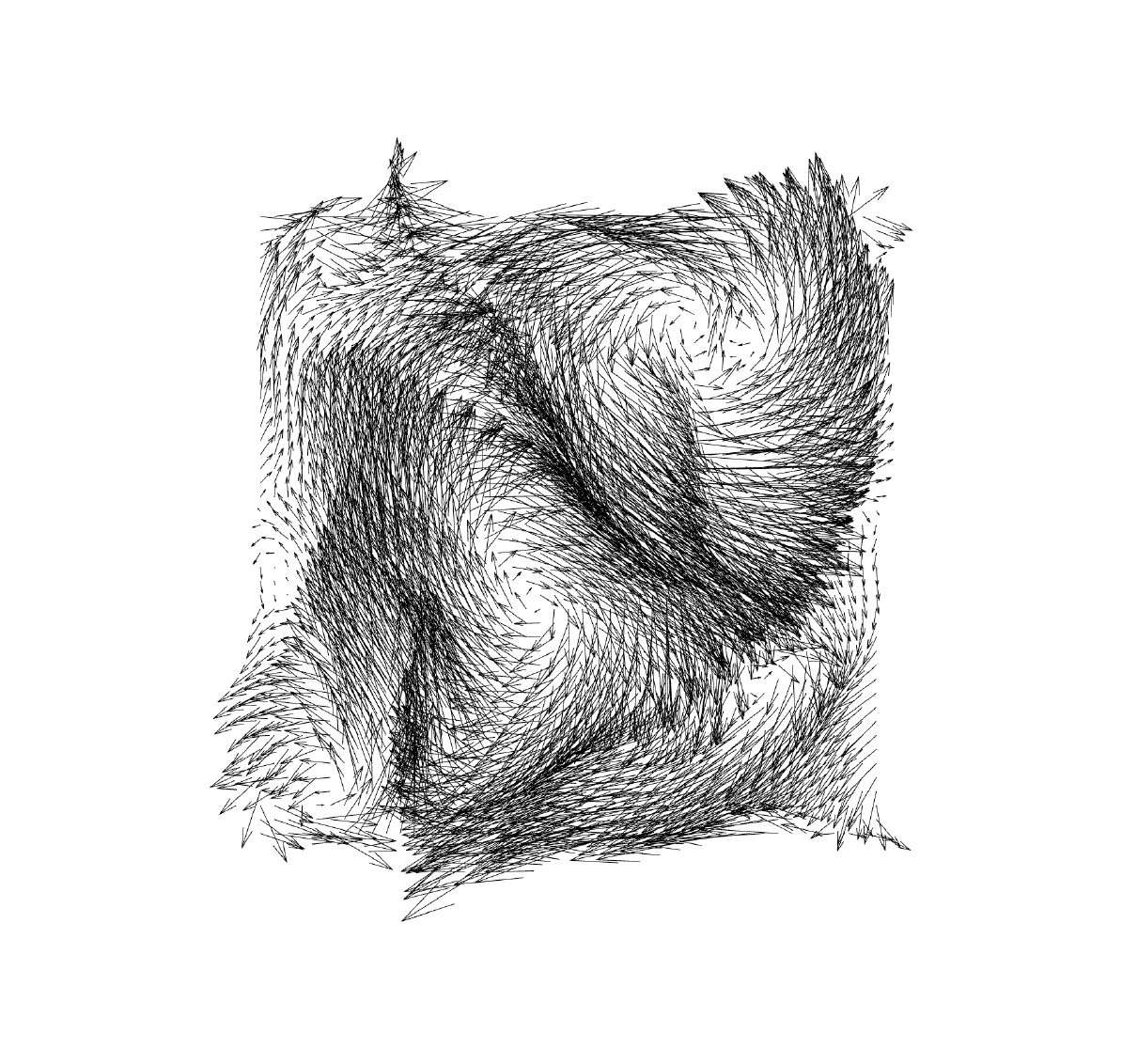}
        \caption{Observed velocity field at a snapshot.} 
        \label{fig:real_velocity_1}
    \end{subfigure}
    \hfill
    \begin{subfigure}[t]{0.485\textwidth}
        \centering
        \includegraphics[width=\linewidth]{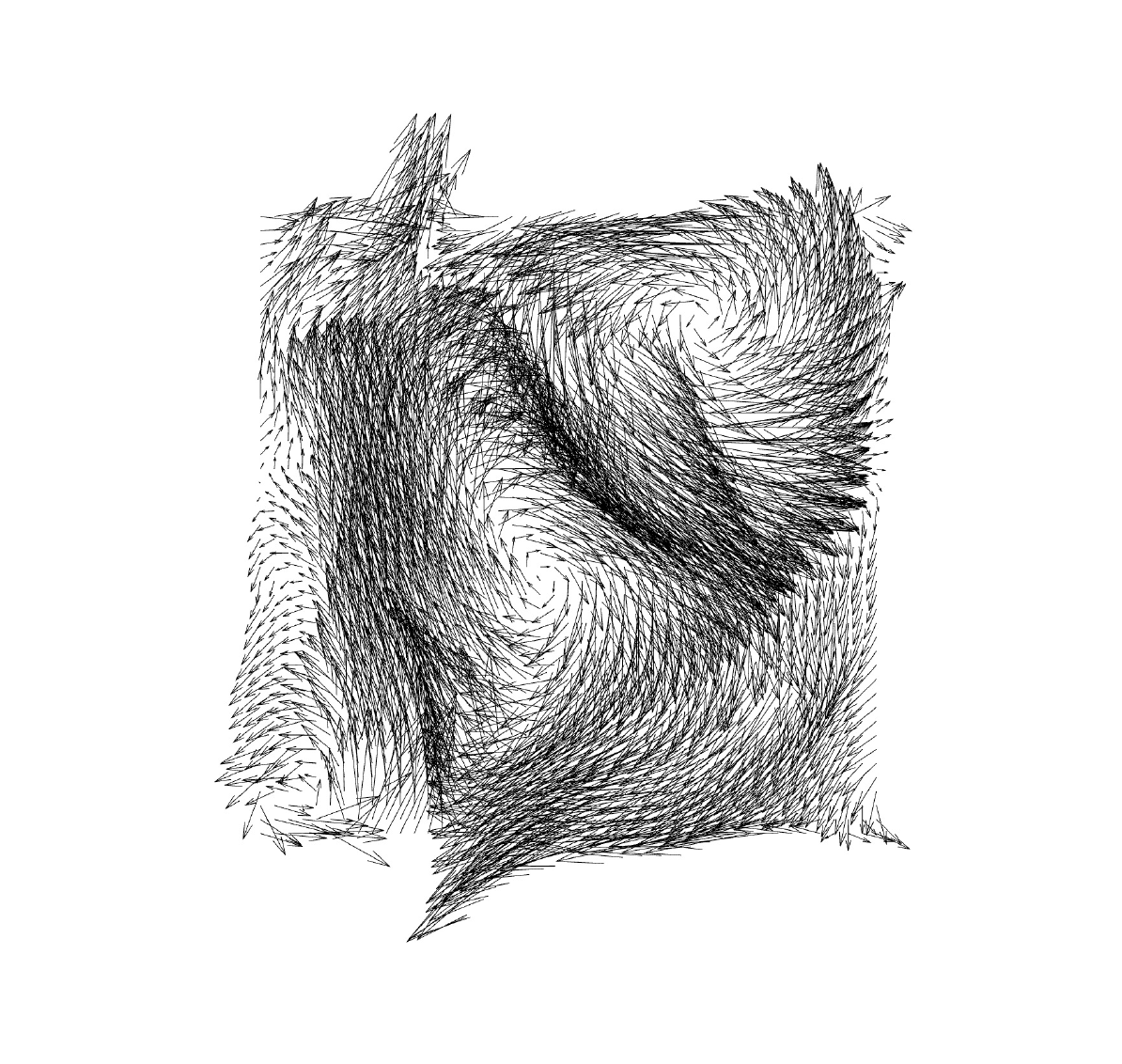}
        \caption{Observed velocity field after 15 min.} 
        \label{fig:real_velocity_2}
    \end{subfigure}

    \vspace{0.6em}

    \begin{subfigure}[t]{0.485\textwidth}
        \centering
        \includegraphics[width=\linewidth]{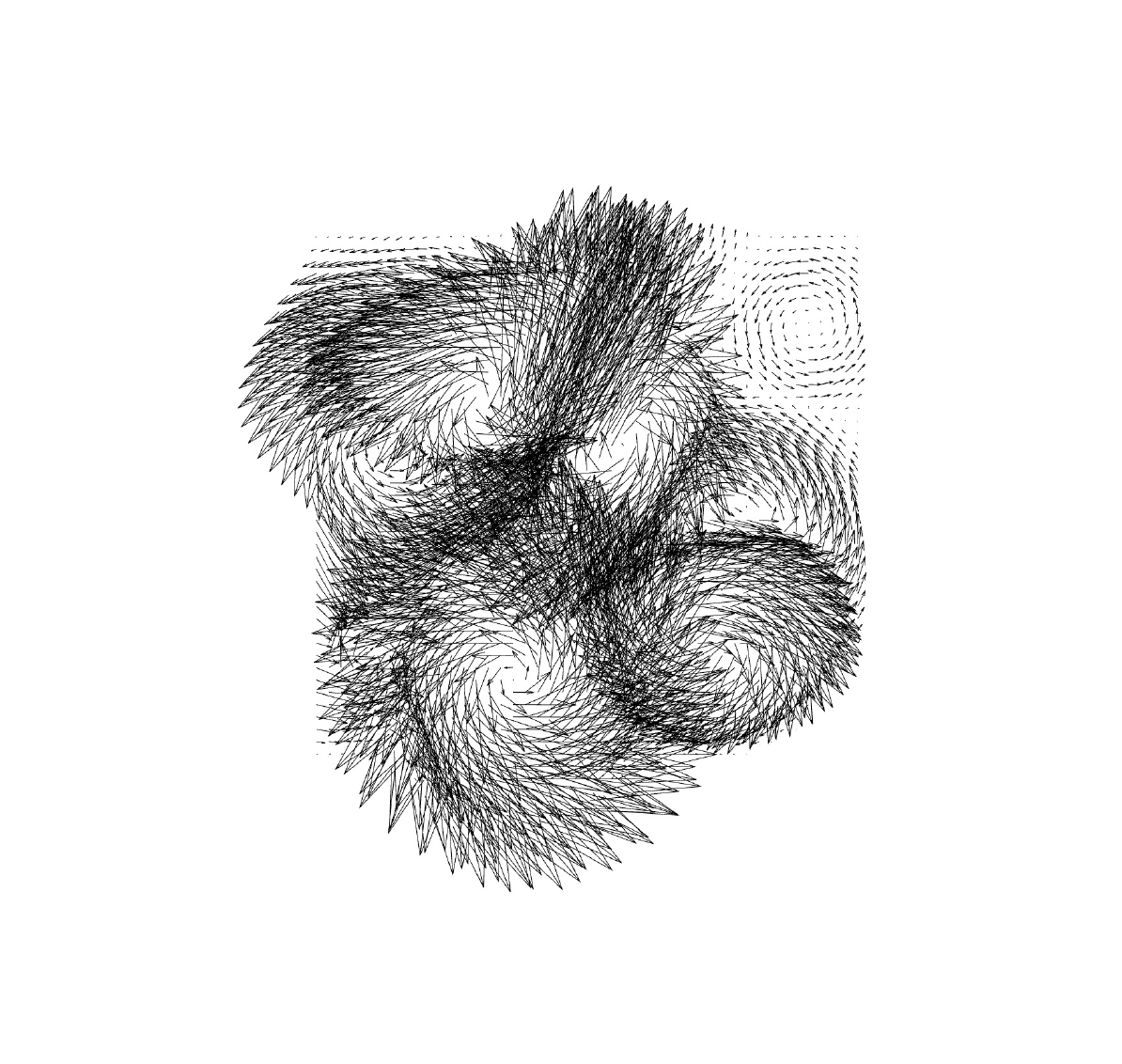}
        \caption{Hawkes-reconstructed velocity field.} 
        \label{fig:sim_velocity_1}
    \end{subfigure}
    \hfill
    \begin{subfigure}[t]{0.485\textwidth}
        \centering
        \includegraphics[width=\linewidth]{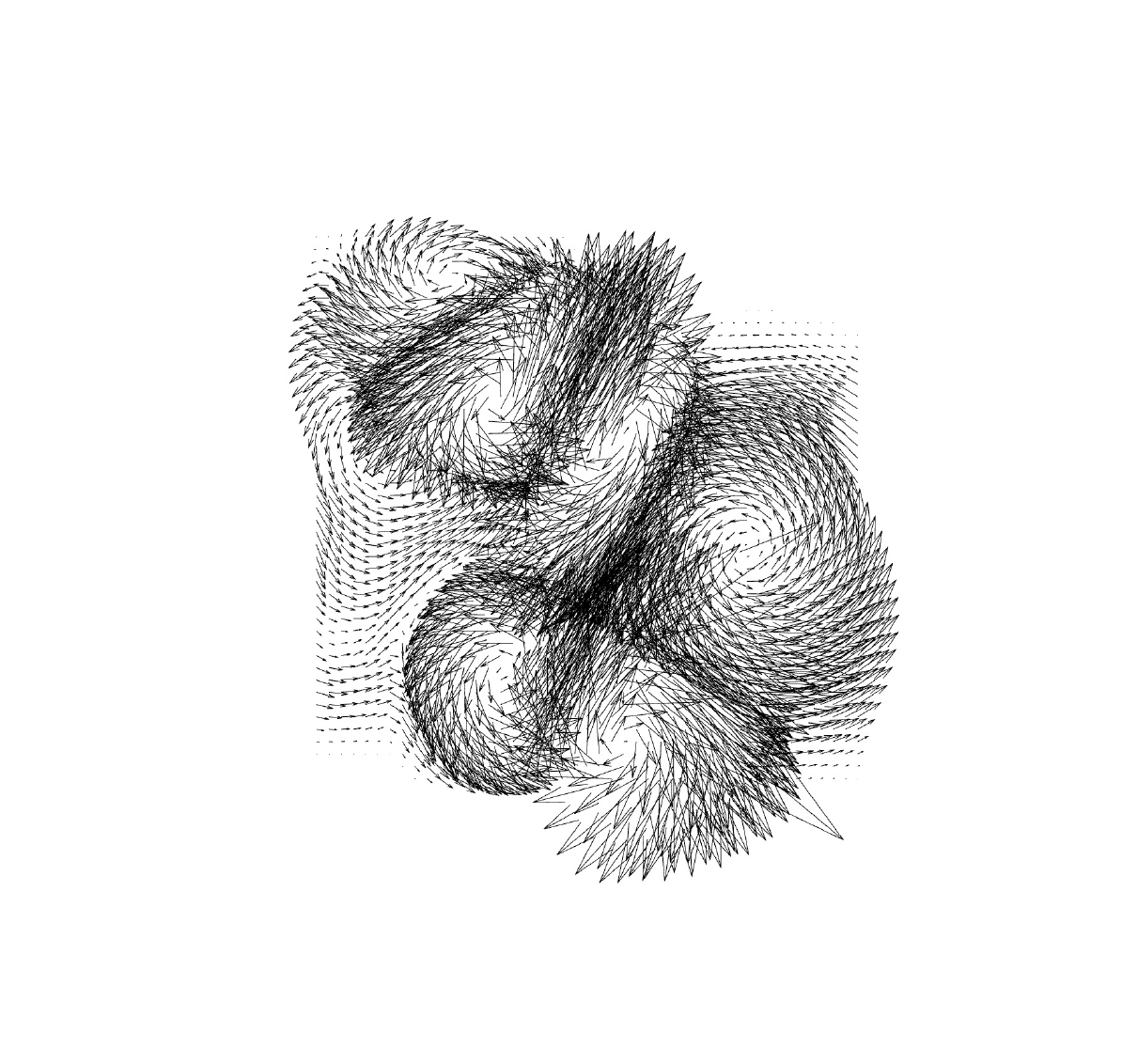}
        \caption{Reconstructed velocity field after 15 min.}
        
        \label{fig:sim_velocity_2}
    \end{subfigure}

    \caption{
        Comparison of observed and reconstructed surface velocity fields.
        Panels (a) and (b) show two consecutive 15-minute snapshots plotted
        directly from the VHF-radar data. Panels (c) and (d) show two
        consecutive 15-minute velocity fields reconstructed from a simulated
        Hawkes eddy catalog. The simulated fields
        reproduce coherent rotational structures and interaction regions,
        but are smoother and more geometrically regular because each eddy is
        represented by the compactly supported, axisymmetric basic-eddy
        profile in Eqs.~\eqref{basic}--\eqref{magnitude}. All four panels use
        the same coordinate limits and one plotting unit per velocity unit.
    }
    \label{fig:observed_simulated_velocity}
\end{figure}

The complete procedure for converting the simulated eddy catalog into
successive velocity-field snapshots is summarized in
Algorithm~\ref{alg:simulated_velocity_reconstruction} in Appendix. For the comparison
below, the algorithm is evaluated on the same two-grid-unit spacing used by
the processed radar velocity fields.
Fig.~\ref{fig:observed_simulated_velocity} compares two consecutive velocity
fields obtained directly from the radar data with two consecutive fields
reconstructed from a simulated Hawkes catalog. The observed snapshots,
shown in Figures~\ref{fig:real_velocity_1} and
\ref{fig:real_velocity_2}, exhibit substantial temporal persistence over the
15-minute interval. The principal circulation regions remain visible in both
snapshots, including a prominent rotational structure in the interior of the
domain, a second circulation region toward the upper-right portion, and an
elongated region of relatively strong flow extending diagonally across the
domain. At the same time, the measured field contains considerable spatial
irregularity: the circulation cells are asymmetric, their boundaries are
distorted, and the velocity directions vary on comparatively small spatial
scales. Such features may reflect eddy deformation, interactions with the
background Florida Current, boundary effects, observational variability, and
flow components that cannot be represented by isolated axisymmetric eddies.

The reconstructed fields in Figures~\ref{fig:sim_velocity_1} and
\ref{fig:sim_velocity_2} reproduce several important qualitative
characteristics of the observations. In particular, they contain multiple
coherent rotational structures, spatially localized circulation centers, and
elongated high-velocity regions produced by the overlap of neighboring
eddies. The persistence of some structures between consecutive simulated
snapshots also reflects the finite lifetimes assigned to the eddies but not the exponential decay of \c{C}inlar flow. Changes
between the two fields are produced by the linear decay of the surviving
eddies together with the appearance or disappearance of eddies according to
their simulated birth times and lifetimes.
Nevertheless, the reconstructed fields are visibly smoother and more
geometrically regular than the observed fields. This is an expected
consequence of constructing the total field by linear superposition. Individual simulated
eddies therefore tend to generate nearly circular circulation patterns,
whereas observed eddies can be stretched, displaced, or deformed by the
surrounding current. Moreover, the present reconstruction does not explicitly
include a large-scale background velocity field, spatially varying eddy
shapes, or the advection of an eddy center during its lifetime. Consequently,
the reconstruction should be interpreted as a stochastic representation of
the eddy-induced component of the flow rather than a pointwise reproduction
of the complete observed velocity field.

The visual comparison
in Fig.~\ref{fig:observed_simulated_velocity} suggests that the model
captures the presence and interaction of coherent rotational structures,
while the greater regularity of the simulated fields indicates directions
for further refinement of the spatial eddy model.

\section{Time-Varying Background Intensity}
\label{6}
The constant-background assumption for the Hawkes model attributes all large-scale changes in the eddy
formation rate either to one global background intensity  or to the triggering kernel. In this section, we  replace the scalar $\lambda_0$ by a time-varying background
rate $\lambda_0(t)$ while preserving  Hawkes
triggering structure to capture the variability observed in Fig.~\ref{fig:constant_background_intensity_comparison}.
Since both 14-day periods show visible nonstationary trends, we  
fit a model in which the background intensity is piecewise constant in time.

\subsection{Modified EM Algorithm}
Let \(I_b=[\tau_{b-1},\tau_b)\), \(b=1,\ldots,B\), be the time bins. The
background rate is modeled as
\[
\lambda_0(t)=\sum_{b=1}^{B}\lambda_{0,b}\mathds{1}_{\{t\in I_b\}}.
\]
For an observed eddy \(i\), let \(b(i)\) denote the bin containing its
arrival time. The E-step background probability becomes
\[
q_i^{(k)}
=
\frac{\lambda_{0,b(i)}^{(k)}}{
\lambda_{0,b(i)}^{(k)}+\sum_{j:t_j<t_i} g(t_i, z_i \mid t_j, z_j,  a_j, b_j, l_j, \Theta^{(k)})}.
\]
Thus the only structural change in the EM algorithm is the background term.
The triggering probabilities \(p_{ij}^{(k)}\) are still normalized by the same
total intensity. The updates for \(K_0,\omega,\rho\) retain the same algebraic form, with the probabilities \(q_i^{(k)}\) and \(p_{ij}^{(k)}\) evaluated using the bin-specific background rate \(\lambda_{0,b(i)}^{(k)}\). The background M-step is now binwise:
\[
\hat{\lambda}_{0,b}^{(k+1)}
=
\frac{\sum_{i=1}^{N}q_i^{(k)}\mathds{1}_{\{t_i\in I_b\}}}
{|S|\,|I_b|},
\qquad b=1,\ldots,B.
\]
This update estimates one exogenous rate for each time bin rather
than a single  \(\lambda_0\).

\subsection{Updated Volterra Solution and Simulation}
The expected total intensity is obtained by replacing the constant background rate in the Volterra equation by the time-varying background intensity. With
\(\tau=t-s\), define
\[
K(\tau)
=
K_0\,\mathbb{E}\!\left[\frac{|a|}{b}\right]\bar{H}_S\,
\omega e^{-\omega \tau}\bar{\gamma}(\tau),
\]
where \(\bar{\gamma}(\tau)=\mathbb{P}(L>\tau)\) is the empirical lifetime
survival function. The spatially integrated expected rate then satisfies
\[
Q(t)
=
|S|\lambda_0(t)
+
\int_0^t K(t-s)Q(s)\,ds. 
\]
We evaluate \(Q(t)\) on the regular grid
\(t_n=n\Delta\). Let \(F_n=|S|\lambda_0(t_n)\). Applying the composite
trapezoidal rule gives  
\[
Q_n
=
F_n
+
\Delta\left[
\frac{1}{2}K(t_n)Q_0
+
\sum_{j=1}^{n-1}K(t_n-t_j)Q_j
+
\frac{1}{2}K(0)Q_n
\right].
\]
Solving explicitly for the current value \(Q_n\) yields the implicit
trapezoidal recursion
\[
Q_n
=
\frac{
F_n+\Delta\left[
\frac{1}{2}K(t_n)Q_0+
\sum_{j=1}^{n-1}K(t_n-t_j)Q_j
\right]
}{
1-\frac{\Delta}{2}K(0)
},
\]
where \(Q_0=F_0\). Note that the background intensity \(F_n\) is constant within each
background bin, whereas \(Q_n\) can vary within and across bins.

In Table \ref{tab:timevarying_bin_results}, the mean number of background eddy arrivals  and branching ratio are summarized. These are estimated with the  the probabilities given in \eqref{optimalprob}.   Specifically, let $q_i^{*}$, $i=1,\ldots, N$, denote the final values obtained  at the end of the  EM algorithm. The mean background arrival rate per unit time over whole domain is computed as $\frac{1} T \sum_{i=1}^{N} q_i^{*}$, while the branching ratio $\hat{n}$ is calculated by 
\begin{equation*}
\hat{n} = \frac{1}{N} \sum_{i=1}^{N} \left(1 - q_i^{*}\right) = 1 - \frac{1}{N} \sum_{i=1}^{N} q_i^{*} \, .
\end{equation*}
We see that the background rate of eddy formation is significantly
lower in the first half of the observation window than in the second half. In
contrast, the branching ratio that measures self-excitation is about three to
four times larger in the first 14 days. More explicitly, the first 14-day
period has about \(8.1\%-8.7\%\) triggered events, whereas the last 14-day
period has about \(2.2\%-2.6\%\). These results are close for the 4-hour and
8-hour fits. We do not consider smaller bin widths, which may overfit the
background rate; therefore, an 8-hour bin width remains a reasonable choice
for estimation.

\begin{figure}[H]
    \centering
    \includegraphics[width=\textwidth]{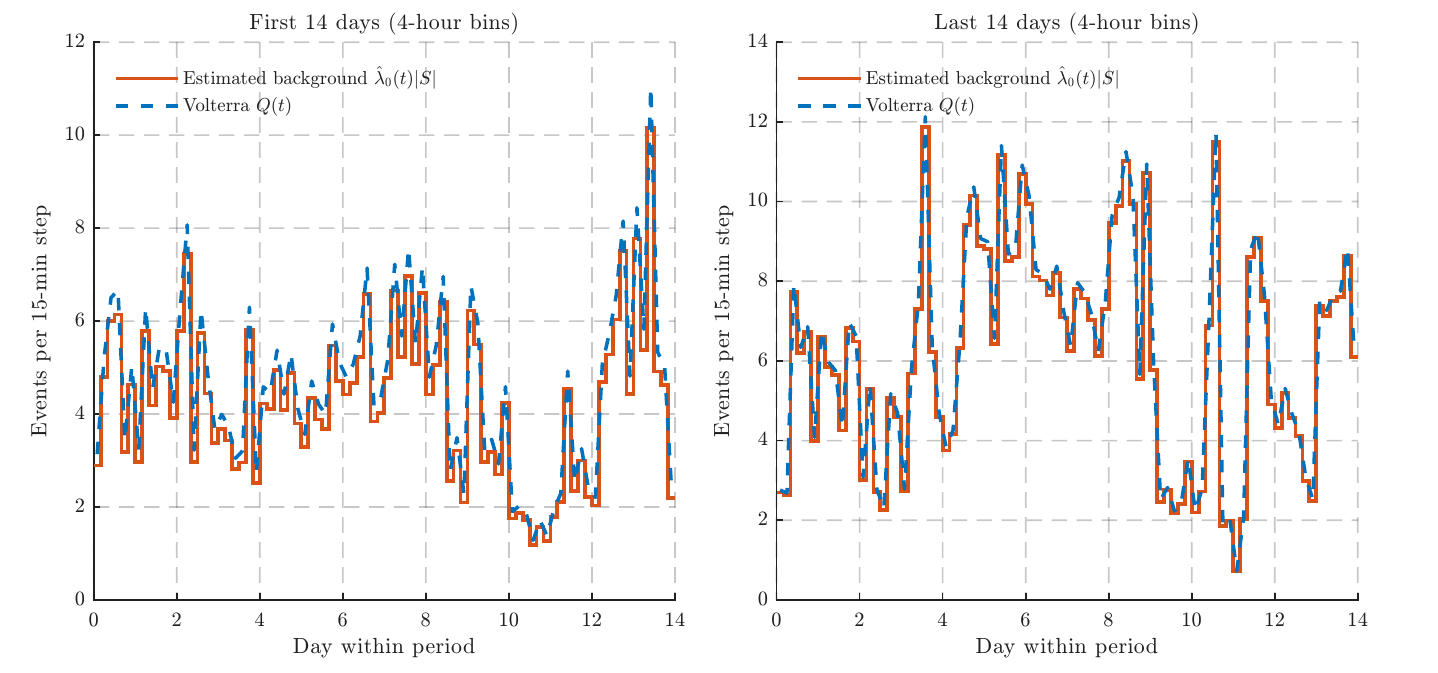}
    \caption{Four-hour bins: estimated background rate
    \(\hat{\lambda}_0(t)|S|\) and Volterra expected rate \(Q(t)\) for the
    first and last 14-day periods.}
    \label{fig:timevarying_background_volterra_4h}
\end{figure}

\begin{figure}[H]
    \centering
    \includegraphics[width=\textwidth]{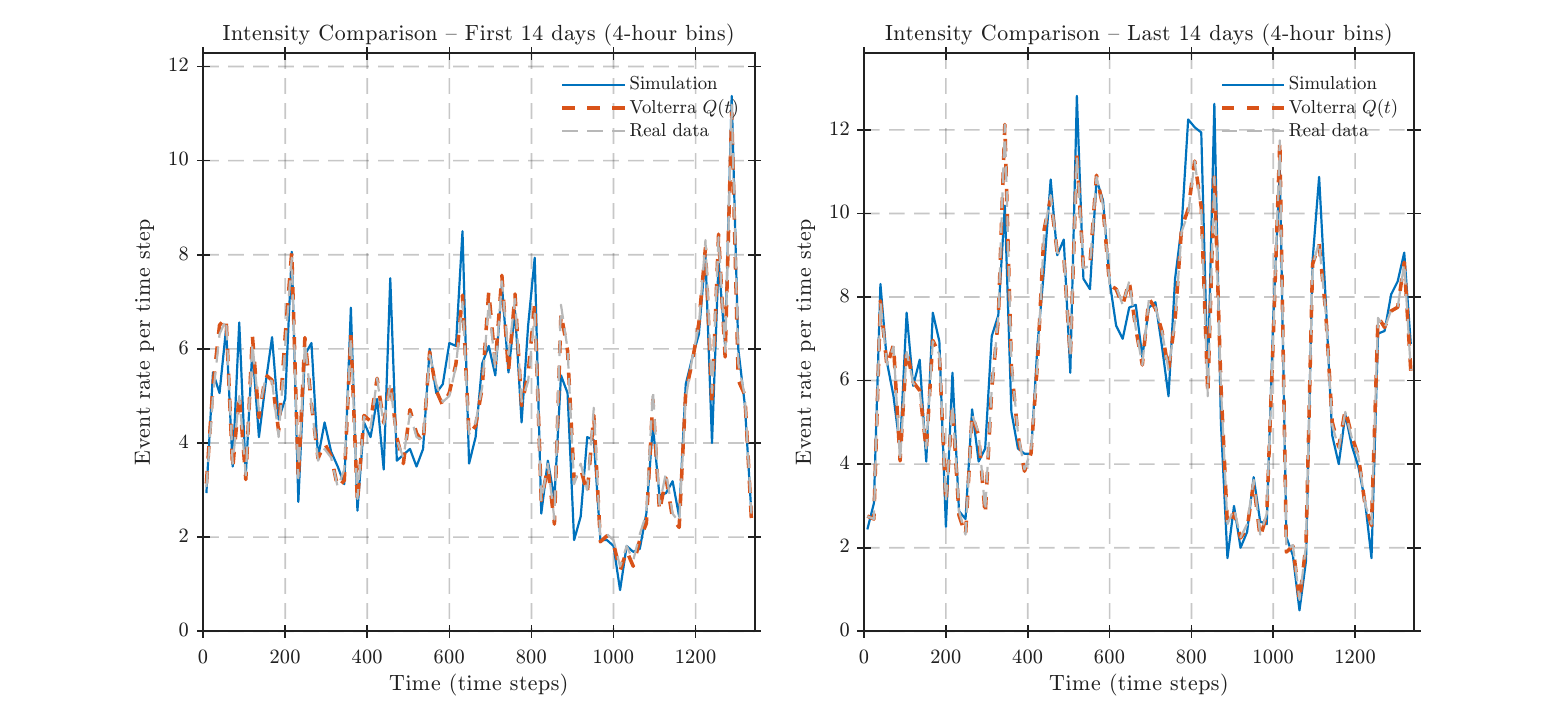}
    \caption{Four-hour bins: intensity comparison showing the empirical rate
    from one representative simulation, the Volterra expected rate \(Q(t)\),
    and the empirical rate from the observed data.}
    \label{fig:timevarying_arrivals_simulation_4h}
\end{figure}

Fig.~\ref{fig:timevarying_background_volterra_4h}   compares the fitted
piecewise-constant background rate $\hat{\lambda}_0(t)|S|$, with the
Volterra expected rate $Q(t)$, using  4-hour  bins. The expected rate 
 $Q(t)$ is very close, but slightly above the
estimated background rate. The small positive difference between these
curves is consistent with the weak but nonzero self-excitation indicated as branching ratio given
in Table~\ref{tab:timevarying_bin_results}, with a more pronounced
triggering contribution during the first 14 days. Fig.~\ref{fig:timevarying_arrivals_simulation_4h} compares the observed
arrival rates  with the expected rate as obtained analytically from Volterra equation and an arbitrary simulated catalog.  The
expected  rate shows the dominant temporal
variation, while the individual simulation and the real data of eddy arrivals 
exhibit stochastic deviations. 

\begin{figure}[H]
    \centering
    \includegraphics[width=\textwidth]{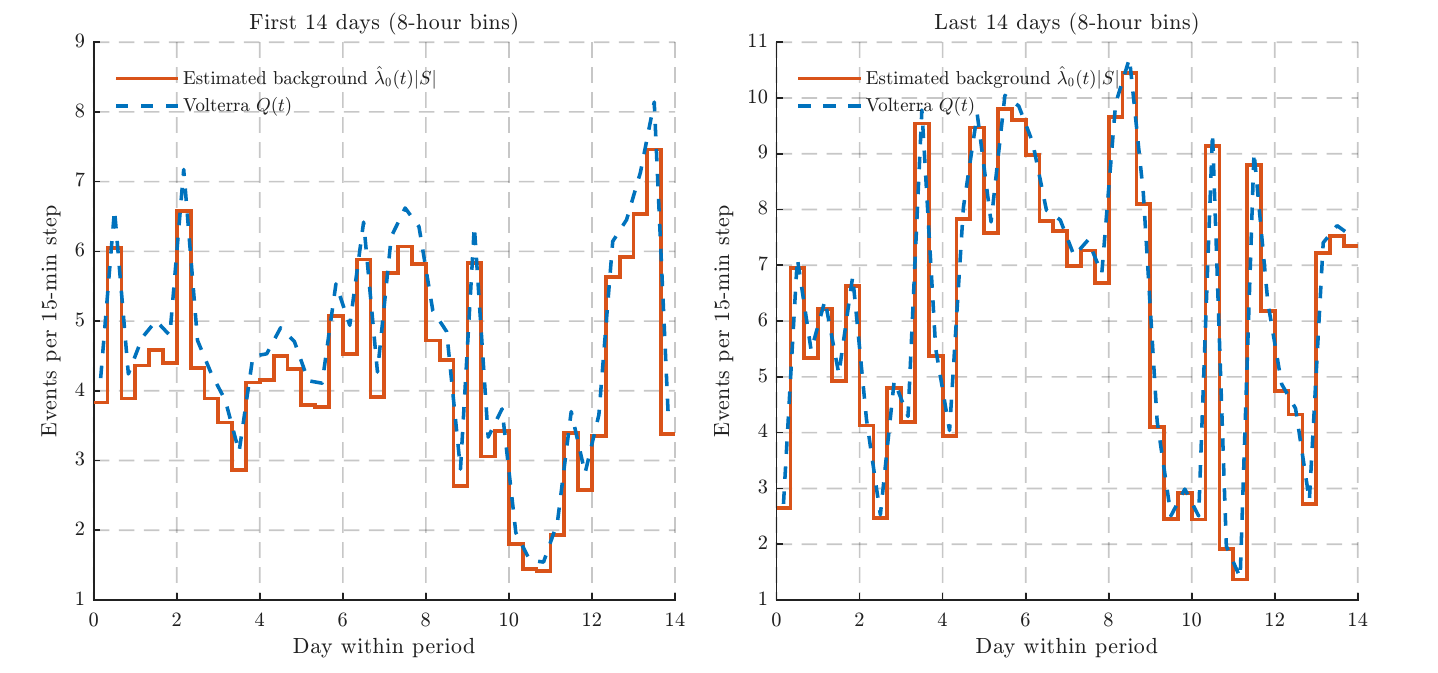}
    \caption{Eight-hour bins: estimated background rate
    \(\hat{\lambda}_0(t)|S|\) and Volterra expected rate \(Q(t)\) for the
    first and last 14-days.
    }
    \label{fig:timevarying_background_volterra_8h}
\end{figure}

\begin{figure}[H]
    \centering
    \includegraphics[width=\textwidth]{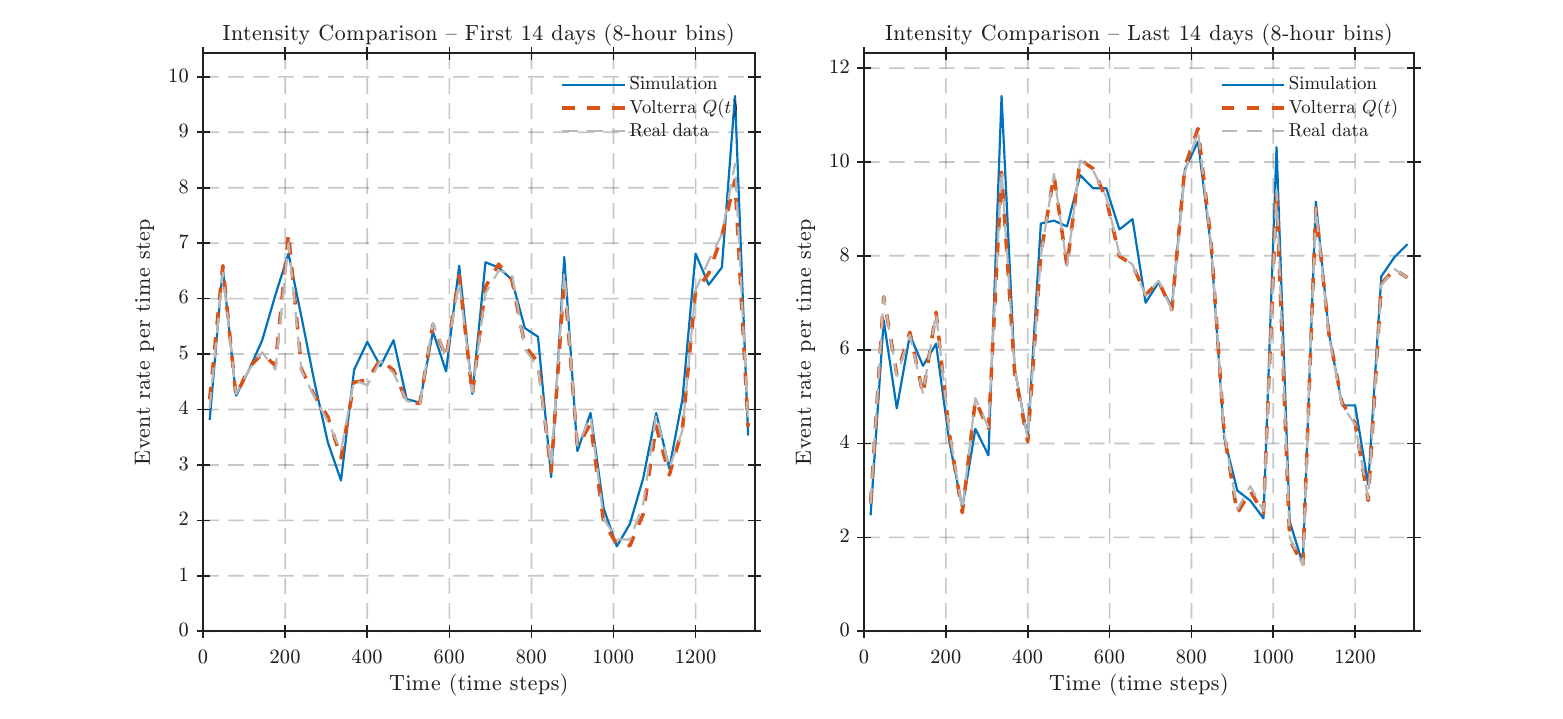}
    \caption{Eight-hour bins: intensity comparison showing the empirical rate
    from one representative simulation, the Volterra expected rate \(Q(t)\),
    and the empirical rate from the observed data.}
    \label{fig:timevarying_arrivals_simulation_8h}
\end{figure}

These comparisons provide an internal consistency check for the EM estimation procedure and the width of the estimation bin for a non-stationary background eddy arrival rate. Therefore, we repeat the estimation and the simulation for 8-hour bins. The 8-hour aggregation
smooths out short timescale fluctuations while preserving the contrast
between the first and last 14-day periods as given in figures \ref{fig:timevarying_background_volterra_8h} and \ref{fig:timevarying_arrivals_simulation_8h}. The coarser time-varying background fits preserve the main qualitative conclusion: most
of the temporal variation is captured by the exogenous background, while the
estimated self-excitation remains weak. The simulated catalogs also track
the observed arrival trends at the 4-hour and 8-hour aggregation scales,
supporting the interpretation that the dominant signal is a time-varying
background rate rather than strong clustering.

\begin{table}[htbp]
\caption{Average of time-varying background fits using 4-hour and 8-hour bins, found from the mean number of background eddy arrivals per 15-minute time step over the full spatial domain.}
\centering
\small
\begin{tabular}{llcc}
\hline
Bin width & Period & Background mean & Branching ratio \\
\hline
\hline
4 hours & First 14 days & 4.285 & 0.0813 \\
4 hours & Last 14 days  & 6.174 & 0.0215 \\
\hline
8 hours & First 14 days & 4.259 & 0.0868 \\
8 hours & Last 14 days  & 6.147 & 0.0258 \\
\hline
\end{tabular}
\label{tab:timevarying_bin_results}
\end{table}

\section{Conclusion}
 \label{7}
  
In this paper, we have introduced a spatio-temporal Hawkes point process  to model submesoscale ocean eddy formation, moving beyond the independence  assumption of  Poisson type velocity fields of Çinlar flow. Using high-frequency radar observations of the Florida Current, the proposed model incorporates both physical flow characteristics such as strain rates and stochastic marks representing eddy amplitude, radius, and lifetime to capture event-to-event excitation dynamics.  
The estimation, simulation, and visualization code is available on 
\href{https://github.com/barisyakar/Eddy-Estimation/}{https:/github.com/barisyakar/Eddy-Estimation/}.  

We have derived a Volterra integral equation governing the expected rate of eddy formation, providing a theoretical mechanism to evaluate long-term, mean-intensity behavior as validation of the computational results. For statistical estimation, we have formulated an Expectation-Maximization (EM) algorithm that treats the latent parent-offspring triggering structure explicitly, guaranteeing numerical stability and reliable parameter convergence. 
For further validation purposes, we have developed a dedicated simulation algorithm for Hawkes type velocity fields. The model is confirmed by demonstrating close agreement between simulated empirical realizations of eddy formation and numerical solutions of the derived Volterra equation.

Our empirical findings reveal that direct self-excitation accounts for only a small fraction of the total intensity of submesoscale eddy formation, as reflected by a small branching ratio $n$. However, despite the modest magnitude of triggering, statistically testing the arrival dynamics confirms that eddy generation significantly deviates from a memoryless Poisson process. Accounting for this subtle self-excitation is critical: it captures localized spatial clustering and temporal bursting, providing a more accurate stochastic parameterization for coastal monitoring and subgrid representations in ocean circulation modeling. 

 \vspace{1cm}

\noindent {\bf Acknowledgements.} This work is supported by TUBITAK 1001 Project 124F343 and Barış Samed Yakar is funded by TUBITAK BIDEB 2211/E – National PhD Scholarship Program.

\section*{Appendix}

\begin{algorithm}[H]
\caption{Expectation-Maximization for Spatio-Temporal Hawkes Process}
\label{alg:em_hawkes}
\begin{algorithmic}
\Require Event catalog $\mathcal{D} = \{ (t_i, z_i, a_i, b_i, l_i) \}_{i=1}^N$, observation windows $S, T$, convergence tolerance $\epsilon$
\Ensure Estimated parameter vector $\hat{\Theta} = (\hat{\lambda}_0, \hat{K}_0, \hat{\omega}, \hat{\rho})$

\State Initialize parameters $\Theta^{(0)} = (\lambda_0^{(0)}, K_0^{(0)}, \omega^{(0)}, \rho^{(0)})$
\State $k \gets 0$
\State $\text{converged} \gets \text{False}$

\While{\textbf{not} converged}
    \vspace{1.5mm}
    \Statex \textbf{// --- E-Step ---}
    \For{$i = 1$ \textbf{to} $N$}
        \State Compute total intensity at event $i$:
        \Statex \qquad $\lambda_i^{(k)} \gets \lambda_0^{(k)} + \sum_{j < i} g(t_i, z_i \mid t_j, z_j, \Theta^{(k)})$
        \State Compute expected background probability:
        \Statex \qquad $q_i^{(k)} \gets \lambda_0^{(k)} / \lambda_i^{(k)}$
        \For{$j = 1$ \textbf{to} $i-1$}
            \State Compute expected triggering probability:
            \Statex \qquad $p_{ij}^{(k)} \gets g(t_i, z_i \mid t_j, z_j, \Theta^{(k)}) / \lambda_i^{(k)}$
        \EndFor
    \EndFor
    \State Compute total expected triggered events: 
    \Statex \quad $\hat{L}^{(k)} \gets \sum_{i=1}^N \sum_{j < i} p_{ij}^{(k)}$
    
    \vspace{1.5mm}
    \Statex \textbf{// --- M-Step ---}
    \State Update background rate: 
    \Statex \quad $\lambda_0^{(k+1)} \gets \frac{\sum_{i=1}^N q_i^{(k)}}{|S| \, T}$
    
    \State Update temporal decay parameter: 
    \Statex \quad $\omega^{(k+1)} \gets \frac{\hat{L}^{(k)}}{\sum_{i=1}^N \sum_{j < i} p_{ij}^{(k)} (t_i - t_j)}$
    
    \State Update spatial decay parameter: 
    \Statex \quad $\rho^{(k+1)} \gets \frac{\hat{L}^{(k)}}{\sum_{i=1}^N \sum_{j < i} p_{ij}^{(k)} \|z_i - z_j\|^2}$
    
    \State Update productivity parameter (conditional on $\omega^{(k+1)}$): 
    \Statex \quad $K_0^{(k+1)} \gets \frac{\hat{L}^{(k)}}{\sum_{i=1}^N \frac{|a_i|}{b_i} \left(1 - e^{-\omega^{(k+1)} l_i}\right)}$
    
    \vspace{1.5mm}
    \Statex \textbf{// --- Convergence Check ---}
    \State $\Theta^{(k+1)} \gets (\lambda_0^{(k+1)}, K_0^{(k+1)}, \omega^{(k+1)}, \rho^{(k+1)})$
    \If{$\|\Theta^{(k+1)} - \Theta^{(k)}\| < \epsilon$} 
        \State $\text{converged} \gets \text{True}$
    \EndIf
    \State $k \gets k + 1$
\EndWhile

\State \Return $\hat{\Theta} \gets \Theta^{(k)}$
\end{algorithmic}
\end{algorithm}

\begin{algorithm}[H]
\caption{Simulation via Branching Process (Cluster Representation)}
\label{alg:simulation_hawkes_cluster}
\begin{algorithmic}
\Require Parameter vector $\Theta = (\lambda_0, K_0, \omega, \rho)$, spatial domain $S$ with area $|S|$, time window $T$, learned spatial density $\hat{f}$
\Ensure Synthetic event catalog $\mathcal{D}_{\text{sim}}$

\vspace{1.5mm}
\Statex \textbf{// --- 0. Offline: Learn Spatial Background ---}
\State Compute $\hat{f}(x,y)$ from empirical eddy locations via 2D Gaussian KDE (Eq.~\ref{eq:spatial_kde})

\vspace{1.5mm}
\Statex \textbf{// --- 1. Background Events (Generation 0) ---}
\State  Compute total background rate: $\mu \gets \lambda_0 \cdot |S|$
\State Draw total background events $N_{\text{b}} \sim \text{Poisson}(\lambda_0 \cdot |S| \cdot T)$
\State Initialize $G^{(0)} \gets \emptyset$
\For{$k = 1$ to $N_{\text{b}}$}
    \State Draw time $t \sim \text{Uniform}(0, T)$
    \State Draw spatial location $(x, y) \sim \hat{f}$ \quad (spatially informed sampling)
    \State Sample features $(a, b, l)$ from empirical distributions
    \State Append $(t, z, a, b, l, \text{parent}=\emptyset)$ to $G^{(0)}$
\EndFor

\vspace{1.5mm}
\State Initialize generation counter $l \gets 0$
\State Initialize final catalog $\mathcal{D}_{\text{sim}} \gets \emptyset$

\vspace{1.5mm}
\Statex \textbf{// --- 2. Offspring Generation (Branching) ---}
\While{$G^{(l)}$ is not empty}
    \State $\mathcal{D}_{\text{sim}} \gets \mathcal{D}_{\text{sim}} \cup G^{(l)}$
    \State Initialize next generation $G^{(l+1)} \gets \emptyset$
    
    \For{\textbf{each} parent event $i \in G^{(l)}$}
        \State Expected offspring: $m_i \gets K_0 \frac{|a_i|}{b_i} \left(1 - e^{-\omega l_i}\right)$
        \State Draw number of offspring $N_i \sim \text{Poisson}(m_i)$
        
        \For{$j = 1$ to $N_i$}
            \State Draw delay $\tau \sim \text{TruncatedExponential}(\text{rate}=\omega, \text{bound}=l_i)$
            \State Offspring time: $t_{\text{new}} \gets t_i + \tau$
            
            \If{$t_{\text{new}} < T$}
                \State Draw offspring location: $z_{\text{new}} \sim \mathcal{N}\left(z_i, \frac{1}{2\rho}\mathbf{I}_2\right)$
                \State Sample features $(a, b, l)$ from empirical distributions
                \State Append $(t_{\text{new}}, z_{\text{new}}, a, b, l, \text{parent}=i)$ to $G^{(l+1)}$
            \EndIf
        \EndFor
    \EndFor
    \State $l \gets l + 1$
\EndWhile

\vspace{1.5mm}
\State \textbf{Sort} $\mathcal{D}_{\text{sim}}$ chronologically by event time $t$
\State \Return $\mathcal{D}_{\text{sim}}$
\end{algorithmic}
\end{algorithm}

\begin{algorithm}[H]
\caption{Velocity-Field Reconstruction from a Simulated Eddy Catalog}
\label{alg:simulated_velocity_reconstruction}
\begin{algorithmic}
\Require Simulated catalog
$\mathcal{D}_{\mathrm{sim}}
=\{(t_i,l_i,x_i,y_i,a_i,b_i)\}_{i=1}^{N}$,
snapshot times $\mathcal{T}$, and spatial grid
$\mathcal{G}=\{(x_m,y_n)\}$
\Ensure Reconstructed velocity fields
$\{U(x_m,y_n;T),V(x_m,y_n;T):T\in\mathcal{T}\}$

\For{\textbf{each} snapshot time $T\in\mathcal{T}$}
    \State Initialize $U(x_m,y_n;T)\gets 0$ and
    $V(x_m,y_n;T)\gets 0$ for all $(x_m,y_n)\in\mathcal{G}$

    \For{\textbf{each} eddy $i\in\mathcal{D}_{\mathrm{sim}}$}
        \If{$t_i\leq T<t_i+l_i$}
            \State Compute the eddy age:
            $\Delta t_i\gets T-t_i$
            \State Apply linear amplitude decay:
            $a_i(T)\gets a_i\left(1-\Delta t_i/l_i\right)$

            \For{\textbf{each} grid point $(x_m,y_n)\in\mathcal{G}$}
                \State $\xi_1\gets(x_m-x_i)/b_i$,
                $\xi_2\gets(y_n-y_i)/b_i$,
                $\rho\gets\sqrt{\xi_1^2+\xi_2^2}$
                \If{$0<\rho\leq 1$}
                    \State $m(\rho)\gets
                    \left[1-\cos(2\pi\rho)\right]/2$
                    \State $u_i\gets
                    -a_i(T)\,[m(\rho)/\rho]\,\xi_2$
                    \State $v_i\gets
                    a_i(T)\,[m(\rho)/\rho]\,\xi_1$
                    \State $U(x_m,y_n;T)\gets
                    U(x_m,y_n;T)+u_i$
                    \State $V(x_m,y_n;T)\gets
                    V(x_m,y_n;T)+v_i$
                \EndIf
            \EndFor
        \EndIf
    \EndFor

    \State Store the snapshot as rows
    $(x_m,y_n,U(x_m,y_n;T),V(x_m,y_n;T))$
\EndFor
\State \Return all reconstructed velocity-field snapshots
\end{algorithmic}
\end{algorithm}

\end{document}